\documentclass{aa}  

\usepackage{graphicx}
\usepackage{xcolor}  % Pacchetto per i colori
\usepackage{comment}

\usepackage{txfonts}
\usepackage[]{hyperref}
\usepackage{natbib}
\bibpunct{(}{)}{;}{a}{}{,} % to follow the A&A style
\begin{document}

   \title{Dust characterization of protoplanetary disks: a guide to multi-wavelength analyses and accurate dust mass measurements}
   \titlerunning{Dust characterization of protoplanetary disks: a guide to multi-wavelength analyses}
\author{Elena M. Viscardi\inst{1}
          \and
          Enrique Macías\inst{1}
          \and 
          Francesco Zagaria\inst{2}
          \and
          Anibal Sierra\inst{3, 4, 5}
          \and 
          Haochang Jiang\inst{1,6}
          \and
          Tomohiro Yoshida\inst{7}
          \and
          Pietro Curone\inst{3}
          }

\institute{European Southern Observatory (ESO), Karl-Schwarzschild-Str. 2, Garching bei München, Germany \\
              \email{Elena.Viscardi@eso.org} 
\and 
    Institute of Astronomy, University of Cambridge, Madingley Road, Cambridge CB3 OHA, UK
\and
    Departamento de Astronomía, Universidad de Chile, Camino El Observatorio 1515, Las Condes, Santiago, Chile
\and 
    Instituto de Astrofísica, Pontificia Universidad Católica de Chile, Av. Vicuña Mackenna 4860, 7820436 Macul, Santiago, Chile
\and
    Mullard Space Science Laboratory, University College London, Holmbury St Mary, Dorking, Surrey RH5 6NT, UK
\and 
    Max Planck Institute for Astronomy, Königstuhl 17, 69117 Heidelberg, Germany 
\and
    National Astronomical Observatory of Japan, 2-21-1 Osawa, Mitaka, Tokyo 181-8588, Japan\\
}

   \date{Received 08/11/2024; accepted 22/01/2025}

% \abstract{}{}{}{}{} 
% 5 {} token are mandatory
 %Measuring dust masses and grain sizes of protoplanetary disks is key to gain a deeper insight into planet formation. These properties can be accurately probed with high-resolution multi-band continuum observations at millimeter wavelengths. With current advance radio-telescopes , which, unfortunately, require large amount of telescope time.
  \abstract
  % context heading (optional)
   {Multi-wavelength dust continuum observations of protoplanetary disks are essential for accurately measuring two key ingredients of planets formation theories: the dust mass and grain size. Unfortunately, they are also extremely time-expensive. }  
   {We aim to investigate the most economic way of performing this analysis by identifying the optimal combination of multi-band observations and the angular resolution that provide accurate results.}
  % aims heading (mandatory)
   {We benchmark the dust characterization analysis on multi-wavelength observations of a compact disk model with shallow rings, and an extended double-ringed disk model. We test three different combinations of bands (in the 0.45 mm $\to$ 7.46 mm range) to see how optically thick and thin observations aid the reconstruction of the dust properties for different morphologies and in three different dust mass regimes. We also test different spatial resolutions (0.05"; 0.1"; 0.2").
   }
  % methods heading (mandatory)
   {Dust properties are robustly measured in a multi-band analysis if optically thin observations are included. For typical disks, this requires wavelengths longer than 3 mm. Instead, fully optically thick observations are not enough to robustly constrain the dust properties. High-resolution (< 0.03"-0.05") is fundamental to resolve the changes in dust content of substructures. However, lower-resolution results still provide an accurate measurement of the total dust mass and of the level of grain growth of rings. Additionally, we propose a new approach that successfully combines lower and higher resolution observations in the multi-wavelength analysis without losing spatial information. We also test individually enhancing the resolution of each radial intensity profile with \texttt{Frank} but we note the presence of artifacts. Finally, we discuss on the total dust mass that we derive from the SED analyses and compare it with the traditional method of deriving dust masses from millimeter fluxes. Accurate dust mass measurements from the SED analysis can be derived by including optically thin tracers. On the other hand, single-wavelength flux-based masses are always underestimated. For the 0.87 mm flux, the underestimation can be more than one order of magnitude. }
  % results heading (mandatory)
   {}


   \maketitle
%
%________________________________________________________________

\section{Introduction}
The dust content of protoplanetary disks sets the most basic initial conditions for planet formation. The total dust mass in disks represents the budget available to form planetesimals, rocky planets, and the cores of giant planets \citep{drazkowska2023planetformationtheoryera}. Its evolution with time is a proxy for the timescales of disk evolution and planet formation. Moreover, the disk mass that can be estimated from the dust mass (by assuming a gas-to-dust ratio) offers insights into which dynamical processes are ongoing during planet-forming stages (e.g., Gravitational Instabilities; \citealt{Longarini_2022}).

Unfortunately, this key ingredient in our planet formation theories is still highly uncertain. The classical mass measurement technique, based on integrated fluxes at (sub-)millimeter wavelengths \citep{Hildebrand}, leads to dust masses of Class II disks that do not reach the solid mass found in exoplanetary systems \citep{manara2018protoplanetary}. While this could suggest that early-stage formation (Class I disks; \citealt{tychoniec2020dust}) and/or late-stage infall are at play \citep{gupta2024tipsy}, new high-resolution observations hint at these values being severely underestimated \citep{liu2022underestimation}. High-resolution observations from facilities like the Atacama Large Millimeter/submillimeter Array (ALMA) or SPHERE at the Very Large Telescope (VLT) (e.g., \citealt{andrews2018disk}, \citealt{garufi2024sphere}) reveal that protoplanetary disks are characterized by various dust substructures \citep{andrews2020observations}, most of which concentrate mm/cm-sized particles. The optical thickness of these dense regions at (sub-)mm wavelengths could be hiding substantially high amounts of dust, leading to low measured dust masses. Although this flux-based mass measurement technique is easy to apply to a large sample of disks, it is unfit for accurately assessing the dust content of protoplanetary disks.

Another key missing piece in planetary formation theory is how mm/cm-sized pebbles grow into km-sized planetesimals. Planetesimal formation via direct growth is severely limited by fragmentation and radial drift \citep{drazkowska2023planetformationtheoryera}. The radial dust redistribution in disks is driven by aerodynamic drag from the surrounding gas, acting like a headwind that slows particles, causing them to drift rapidly toward the disk center \citep{whipple}. Immersed in a turbulent gas medium, the relative velocities of grains increase, eventually leading to fragmentation rather than accretion \citep{2012Birnstiel}.

The ubiquitous detection of substructures in high-resolution ALMA observations suggests a way to ease this rapid depletion of the disks' dust reservoir. If these substructures result from pressure maxima, they can act as dust traps, slowing drifting particles \citep{pinilla2012ring}. In these dense dust regions, the dust’s back-reaction on the gas slows the local gas velocity, reducing the headwind experienced by the dust. For sufficiently high dust-to-gas ratios (\citealt{2021Li&Youdin}, \citealt{2024Lim}), this positive feedback leads to the formation of self-gravitating clumps of particles in the mid-plane that rapidly grow into kilometer-sized planetesimals. This process, known as the streaming instability, is currently the leading theory for planetesimal formation \citep{youdin2005streaming}. By measuring the level of grain growth and dust density in substructures, we can test if the physical criteria required to trigger streaming instabilities are met \citep{2021Scardoni}.

Polarization observations at (sub-)mm wavelengths have been proposed as a method to measure grain sizes in disks. Recent measurements from ALMA suggest sub-millimeter maximum dust sizes \citep{kataoka2017evidence}, in contrast to the (sub-)centimeter maximum particle sizes predicted by simulations of coagulation/fragmentation dust evolution with dust trapping \citep{2012Birnstiel}. However, different interpretations of the polarization pattern and fraction yield differing outcomes that may reconcile these discrepancies (e.g., a double dust population, \citealt{kataoka2017evidence}; particle porosity, \citealt{zhang2023porous}). At present, the complex interpretation of these polarization observations hinder a clear determination of the maximum grain size.

%At the moment, no robust constrain on the particle size distribution in protoplanetary disks can be derived from this analysis. 

Alternatively, multi-wavelength continuum observations of the dust thermal emission at (sub)millimeter wavelengths remain one of the most efficient tools to study both the particle size distribution and the dust surface density in protoplanetary disks. The Spectral Energy Distribution (SED) of the thermal dust emission of disks depends on the dust temperature and, if optically thin detections are included, on the surface density and on the maximum particle size through the frequency dependence of the dust opacity (see Subsection \ref{ssec:radiative}). The few bright sources that have been observed at high resolution at various wavelengths all show evidence of accumulations of pebbles in substructures, in agreement with our current view on planet formation (\citealt{carrasco2019radial}; \citealt{sierra2021molecules}; \citealt{macias2021characterizing}; etc.). To confirm these preliminary findings, more detailed multi-wavelength studies of a larger, unbiased sample of disks are necessary. However, differences in sensitivity and resolution achievable across various frequencies by advanced facilities, such as ALMA and the Jansky Very Large Array (VLA), currently pose significant limitations to this goal.

High-resolution disk observations outside the 211 GHz–373 GHz range (ALMA Bands 6 and 7) are uncommon due to the challenges of achieving both high resolution and sensitivity at lower frequencies, where the emission is fainter (ALMA Bands 3 and 4), and at higher frequencies, where atmospheric conditions are more demanding (ALMA Bands 8, 9, and 10). Moreover, both high-resolution multi-wavelength analyses (e.g., \citealt{carrasco2019radial}; \citealt{macias2021characterizing}), as well as population synthesis models \citep{delussu2024population}) have shown that dense disk regions, such as the inner disk or rings, can remain marginally optically thick even at wavelengths up to 3 mm (ALMA Band 3). Unfortunately, optically thin, longer-wavelength observations currently offer coarser resolution than required to resolve substructures (< 0.03"-0.05"). ALMA Band 1 (6-8.6 mm)  data can achieve a maximum resolution of 0.1". Although the VLA can provide somewhat higher resolutions (0.05"-0.06" at 7-9 mm), its observations are of lower quality compared to ALMA due to the VLA's limited UV coverage and poorer phase stability. Achieving sufficient sensitivity requires extensive observation times (e.g., $\sim$32 hours for the 9 mm VLA image of HL Tau; \citealt{carrasco2019radial}). Additionally, joint ALMA and VLA observations are limited to a small sample of sources in a narrow region of the sky. As a result, only a limited number of SED analyses of disks incorporate both high-resolution optically thick and thin data, which are crucial for accurate measurements of dust properties.

All of the previously mentioned limitations make multi-frequency continuum observations among the most time-consuming methods for studying protoplanetary disks. Thus, it is fundamental to investigate the most economic way of performing this dust characterization analysis.

In this work, we address this question by benchmarking the state-of-the-art dust characterization analysis on simulated multi-wavelength observations of protoplanetary disks. First, we aim to identify the optimal combination of multi-wavelength observations that achieves accurate results while minimizing observational demands on telescope resources. Since the best multi-band setup depends on the optical depth of the observations, we vary the total dust content of our model disks while performing this analysis.   
Additionally, we thoroughly test how the differences in quality and spatial resolution and/or the absence of optically thin observations - which characterize the majority of the available spatially-resolved SED analyses - affect the determination of the dust content of protoplanetary disks. Lastly, we propose an innovative approach that enables the inference of dust properties at high angular resolution using economical, medium-resolution ALMA Band 1 observations. This method not only exploits ALMA’s higher quality to capture even fainter objects compared to VLA, but also provides a more cohesive dataset for multi-band analyses by ensuring a consistent field of view across observations. Together, these advantages allow us to extend dust characterization to a larger sample of disks, enhancing the statistical robustness of our findings.

%How the differences in quality and spatial resolution and/or the absence of optically thin observations - which characterize the majority of the available spatially-resolved SED analyses - affect the determination of the dust content of protoplanetary disks has not been thoroughly tested. In this work, we address this gap. First, we benchmark the state-of-the-art dust characterization analysis on simulated multi-wavelength observations of protoplanetary disks. We then explore possible solutions to overcome its limitations. In particular, we propose an innovative approach that enables to infer dust properties at high angular resolution using high-quality and yet economical medium-resolution ALMA Band 1 observations. This method provides an opportunity to extend dust characterization to a larger sample of disks, enhancing the statistical significance of our findings.

In Section \ref{sec:data} we describe the simulated multi-frequency dust thermal emission of disks. In Section \ref{sec:analysis}, we introduce the state-of-the-art multi-wavelength SED analysis. The results of benchmarking this technique on simulated observations are shown in Section \ref{sec:benchmark}. In particular, in Subsection \ref{ssec:bands}, we discuss how the dust thermal emission at various frequencies aid the reconstruction of the dust properties from an SED analysis. Instead, in Subsection \ref{ssec:beams}, we discuss how the limited resolution of real observations affects the measurement of the dust properties. Finally, in Section \ref{sec:improv}, we propose new approaches to the dust characterization technique that allows to extend this method to a larger, unbiased sample of disks.

%__________________________________________________________________

\section{Simulated multi-wavelength observations}
\label{sec:data}
To quantify the effectiveness of the current SED analysis procedure in constraining dust properties from real observations, we apply this technique to simulated multi-wavelength observations of disks. 

We design realistic observations of the dust thermal emission of a protoplanetary disk by first evolving an interstellar medium (ISM) dust population over 1 Myr. This evolution accounts for the coagulation and fragmentation processes expected within the disk, as well as dust transport (see Subsection \ref{ssec:dustpy}). Then, we analytically compute the multi-band thermal emission of this evolved dust population, assuming that it can be described as a 1D vertically isothermal slab (see Subsection \ref{ssec:radiative}). We include in this analysis the emission at 0.45 mm (ALMA Band 9), 0.87 mm (ALMA Band 7), 1.29 mm (ALMA Band 6), 3.07 mm (ALMA Band 3), and 7.46 mm (ALMA Band 1 or VLA Q Band). We use Band 9 observations for optically thick detections of dust thermal emission, as Band 10 requires very strict weather conditions for observations. While including both Band 7 and 6 might not be necessary since they have similar expected optical depths, the combination of both these highly requested observations mitigates the uncertainties coming from the noise and the flux calibration errors. The emission at both Band 4 and Band 3 is expected to be marginally thin but, because of the slightly lower frequency, Band 3 observations are preferred. Both Band 8 and Band 5 detections are less common than Band 9/7 or Band 6/4. For the long-wavelength observations, we consider a frequency of 40 GHz, which can be observed by both ALMA and the VLA, enabling comparative analysis. Similar results can be obtained with the 9 mm emission observed by the VLA in Ka Band. 

We design two disk models: a compact disk (with a radius after 1 Myr of disk evolution of $R_{68\%} \sim 0.15" = 22$ au and $R_{99\%} \sim 0.39" = 55$ au at 140 pc, in Band 7 continuum emission), with a low contrast and small scale ringed morphology that resembles TW Hya \citep{macias2021characterizing}; and an extended (with a radius after 1 Myr of disk evolution of $R_{68\%} \sim 0.35" = 50$ au and $R_{99\%} \sim 0.82" = 115$ au at 140 pc, in Band 7 continuum emission) double-ringed disk, similar to HD 163296 \citep{guidi2022distribution}. We simulate multi-wavelength radio-interferometric observations of these observations at three different resolutions: 0.05", which represents the maximum resolution currently achievable at 7 mm by the VLA and at 3 mm by ALMA; 0.1", the limit of ALMA Band 1 with ALMA's most extended configuration; and 0.2", a practical resolution for most disks (see Subsection \ref{ssec:simobs}).

\subsection{Simulated dust distributions}
\label{ssec:dustpy}
\begin{figure*}[h]{}
   \centering
   \includegraphics[width=1.\textwidth]{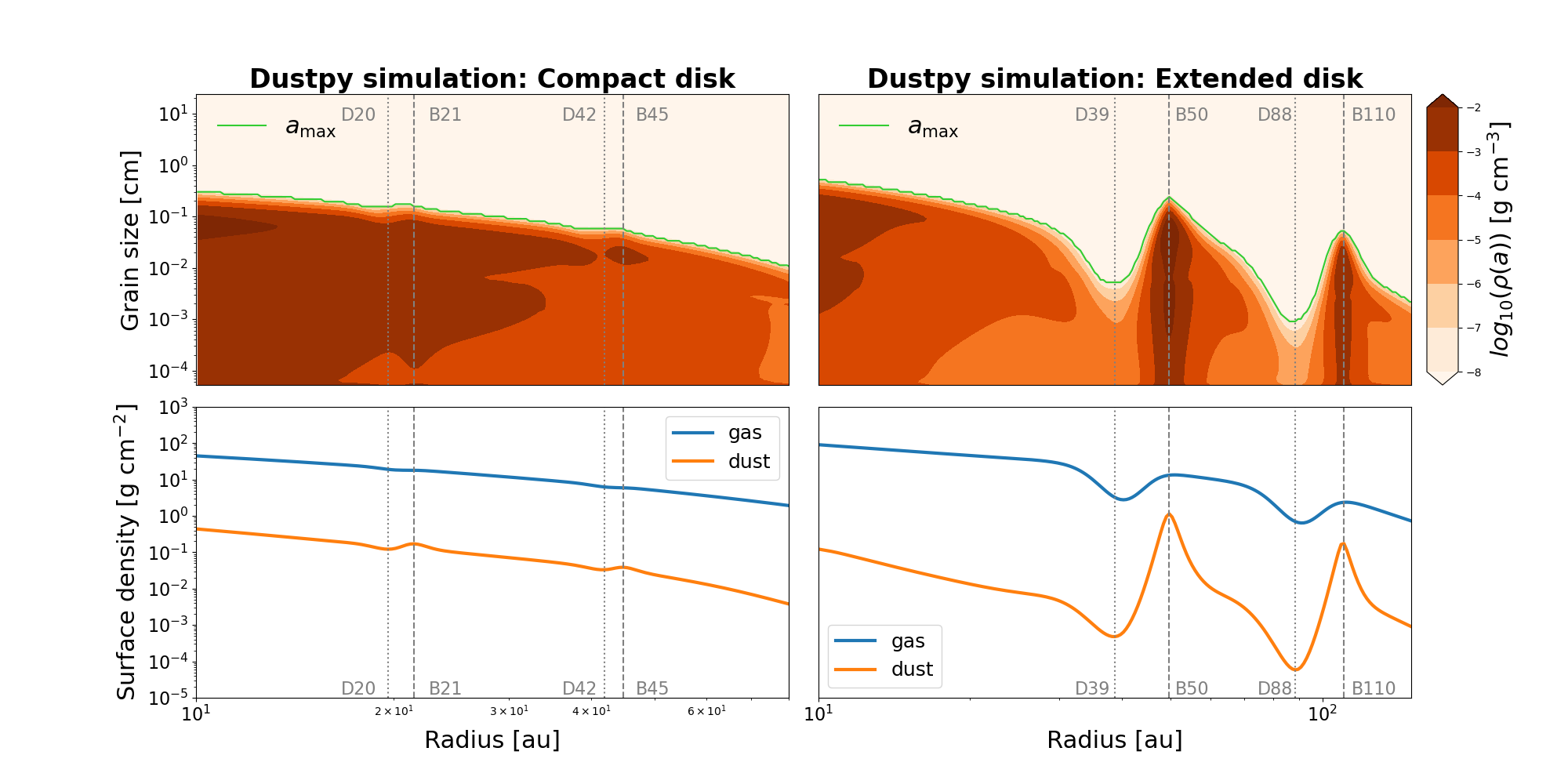}
   \caption{\texttt{DustPy} simulation outputs at t = 1 Myr for the compact disk (\textit{left panel}) and the extended disk models (\textit{right panel}). \textit{Top panel:} Grain size distribution of the dust particles as a function of the radial distance from the star. At each radius, what we identified as the \textit{maximum grain size} that has non-negligible dust density, is highlighted in lime. 
   \textit{Bottom panel:} Radial profiles of gas (\textit{in blue}) and dust (\textit{in orange}) total column densities. The vertical gray lines represent the position of the bright (B) and dark (D) gaps in the dust surface density profile.}
    \label{fig:dustpy}%
\end{figure*}
We employ the open-source code \texttt{DustPy} \citep{Stammler_2022} to simulate the evolution of the dust mass distribution in the two disks with substructures we are modeling. This code solves the Smoluchowski coagulation equation \citep{Smoluchowski} while accounting for dust fragmentation and transport. Following the results of \citet{2024Jiang}, we assume a population of fragile pebbles with a fragmentation velocity of 1 m/s. For the initial grain size distribution, we adopt the default MRN distribution observed by \citet{Mathis1977} in the Interstellar Medium (ISM): $n(a) \, \mathrm{d}a \propto a^{-q} \, \mathrm{d}a $, with $q = 3.5$, a minimum grain size of 0.005 $\mu$m and a maximum grain size of 1 $\mu$m.

The gas surface density evolves with the master equation of viscous accretion \citep{Lynden-Bell}, while the dust surface density evolution is solved by the advection-diffusion equation \citep{birnstiel2011dust}. The relative velocities of dust particles prior to collisional growth accounts for thermal Brownian motion, vertical stirring and settling, turbulent mixing, as well as azimuthal and radial drift.

We initialize the gas surface density as the similarity solution of \citet{Lynden-Bell}:
\begin{equation}
\Sigma_\mathrm{g} = \Sigma_\mathrm{g, 0} \, \Biggl( \frac{r}{r_\mathrm{c}}\Biggr)^{-\gamma} \, \exp \Biggl[ -\Biggl( \frac{r}{r_\mathrm{c} }\Biggr)^{2-\gamma} \Biggr],
\end{equation}
with $\gamma =1$ and a characteristic radius $r_\mathrm{c} = 100$ au. For both models, we assume an initial gas mass of 0.05 $M_\odot$ and an initial gas-to-dust ratio of 100. The disk is assumed to be vertically isothermal. This is a good approximations since we aim to analyze the (sub)millimeter emission of disks, which is tracing the geometrically thin midplane layer of the disk where large dust grains have settled \citep{2016Pinte}. We define the temperature profile as
\begin{equation}
T_\mathrm{dust} = T_\mathrm{0} \, \Biggl(\frac{r}{r_\mathrm{c}} \Biggr)^{-0.45},
\label{eq:Tdust}
\end{equation}
with the typically assumed $T_\mathrm{0} = 20$ K at $r_\mathrm{c} = 100$ au.
To simulate the dust evolution in disks with substructures, we introduce pressure traps by radially modifying the gas viscosity as \citep{Dullemond_2018}:  
\begin{equation}
\alpha_{\nu, r} = \frac{\alpha_{\nu}}{F(r)}, \, \, \, \, \, F(r) = \exp \Biggl[-A\, \exp \Biggl( -\frac{(r-r_0)^2}{2\, w^2}\Biggr) \Biggr],
\end{equation}
with $\alpha_{\nu} = 10^{-4}$. The radial mixing ($\delta_r$), turbulent mixing ($\delta_t$) and vertical mixing ($\delta_v$) parameters are fixed to a value of $10^{-4}$. 
Our extended disk model is characterized by two deep gaps (A = 2.0) at positions $r_0=$ 50 au and $r_0=$ 100 au with a width of $w = $ 1 au. 
On the contrary, to obtain the compact disk model, we introduce two shallow gaps (A = 0.1 au) at half the distance from the star ($r_0=25\,{\rm au}$ and 50 au) and with half the width ($w=0.5\,{\rm au}$).

The results of the \texttt{DustPy} simulations after 1 Myr are shown in Figure \ref{fig:dustpy}. The top panel shows the grain size distribution as a function of the radial distance from the star for the compact disk (\textit{left panel}) and the extended disk (\textit{right panel}). At each radius, what we identify as the \textit{maximum grain size} $a_\mathrm{max}(r)$, i.e. the maximum grain size that has non-negligible dust density, is highlighted in \textit{lime}. Here we define it as the grain size corresponding to a dust mass density in the $10^{-6} - 10^{-8}$ g/cm$^{3}$ range. This range was selected through visual inspection of Figure \ref{fig:dustpy}, as lower densities do not reflect the expected conditions in the midplane of a protoplanetary disk. Note that this choice is somewhat arbitrary but a different choice of threshold would simply move the maximum grain size profile slightly up or down, which does not affect the aim or results of this paper. In the bottom row, the corresponding gas and dust surface densities distributions are shown. The vertical gray lines represent the position of the bright (B) and dark (D) rings in the dust surface density profile.

The maximum grain size in the compact disk is $\sim 3$ mm in the inner disk. It reaches a value of 1.5 mm in the inner ring before decreasing to sub-millimeter sizes in the outer disk. Instead, the maximum grain sizes at the two rings of the extended disk are respectively: $\sim$ 2.5 mm and 0.5 mm. This is in agreement with the results of SED analyses of the handful of T-Tauri stars for which we have high-resolution multi-wavelengths observations (\citealt{sierra2021molecules}; \citealt{carrasco2019radial}; \citealt{tazzari2016multiwavelength}).

\subsection{Analytical Radiative Transfer}
\label{ssec:radiative}
To compute the continuum emission at (sub)mm-wavelengths of the simulated grains population of our model disks, we assume the disks to be azimuthally symmetric, vertically isothermal, and razor-thin. With these assumptions, the dust thermal emission from the disk mid-plane can be computed as a 1D vertically isothermal slab (\citealt{1993Miyake}; \citealt{sierra2019analytical}):
\begin{equation}
\label{eq:inu}
I_\nu = B_\nu(T_\mathrm{dust}) \left[(1 - \mathrm{e}^{-\tau_\nu / \mu}) + \omega_\nu \, F(\tau_\nu, \omega_\nu)\right],
\end{equation}

where $B_\nu(T_\mathrm{dust})$ is the Black-Body emission at the dust temperature $T_\mathrm{dust}$ and frequency $\nu$. $\tau_\nu = \Sigma_\mathrm{dust}\, \chi_\nu$ is the optical depth, defined as the product of the dust surface density $\Sigma_\mathrm{dust}$ and the total dust opacity $\chi_\nu$.

At millimeter wavelengths, emission due to dust scattering is not negligible. On the contrary, dust scattering can considerably reduce the intensity of an optically thick disk \citep{Zhu2019}. Thus, here we consider the total dust absorption $\chi_\nu$ as the sum of the absorption opacity $k_\nu^\mathrm{a}$ and the effective scattering opacity $k_\nu^{\mathrm{s},\mathrm{eff}}$. The effective scattering opacity is an approximation for the anisotropic scattering (\citealt{1941Henyey}; \citealt{Birnstiel2018}) and is related to the scattering opacity ($k_\nu^\mathrm{s}$) as: $k_\nu^{\mathrm{s},\mathrm{eff}} = (1-g_\nu) \, k_\nu^\mathrm{s}$, where $g_\nu$ is the forward scattering parameter. The albedo is defined as $\omega_\nu = k_\nu^\mathrm{s}/ (k_\nu^\mathrm{a}+k_\nu^\mathrm{s})$. Lastly, $\mu = \cos(i)$ is the cosine of the disk inclination angle $i$. Here, we assume face-on ($\mu=1$) disks.

Since scattering is taken into account, the solution for the radiative transfer depends also on

\begin{equation}
\begin{aligned}
\label{eq:scat}
F(\tau_\nu, \omega_\nu) = \frac{1}{\exp{(-\sqrt{3}\epsilon_\nu \,\tau_\nu)\,(\epsilon_\nu-1) - (\epsilon_\nu+1) }} \times \\ 
\Biggl[  \frac{1-\exp(-(\sqrt{3}\epsilon_\nu + 1/\mu)\tau_\nu)}{\sqrt{3}\epsilon_\nu \, \mu + 1} + \frac{\exp(-\tau_\nu/\mu) - \exp(-\sqrt{3}\epsilon_\nu \tau_\nu)}{\sqrt{3}\epsilon_\nu \, \mu - 1}  \Biggr],
\end{aligned}
\end{equation}
where $\epsilon_\nu = \sqrt{1-\omega_\nu}$.

Given a dust temperature radial profile $T_\mathrm{dust}(r)$ and the absorption and scattering opacities $k_\nu^\mathrm{a}$ and $k_\nu^\mathrm{s}$, the continuum intensity in equation \ref{eq:inu} can be computed for the two dust surface density models ($\Sigma_\mathrm{dust}$) at all wavelengths. We define $T_\mathrm{dust}(r)$ as in equation \ref{eq:Tdust}. While this equation does not account for the thermal behavior of gaps and rings, performing a full radiative transfer calculation would be computationally expensive and would require detailed information on the vertical structure of disks, which is not simulated in \texttt{DustPy}. Additionally, a more physical description would not significantly affect the scope of our work.

Since the dust population in disks consists of a continuous size distribution instead of a single size, as highlighted by both real observations \citep{2018Avenhaus} as well as models of grain growth with fragmentation (\citealt{Weidenschilling1984}; \citealt{DullemondDominik2005}; \citealt{Brauer2008}; \citealt{Birnstiel2010}), we average at each radius the size-dependent opacities $k_\nu^\mathrm{a}(a)$ and $k_\nu^\mathrm{s}(a)$ over the particle mass distribution:
\begin{equation}
k_\nu^\mathrm{ a/s, \,tot}(r)= \frac{\int_{a_\mathrm{min}(r)}^{a_\mathrm{max}(r)} n(a) \, m(a) \, k_\nu^\mathrm{ a/s} (a) \,  \mathrm{d}a}{\int_{a_\mathrm{min}(r)}^{a_\mathrm{max}(r)} n(a) \, m(a) \,  \mathrm{d}a}.
\label{eq:ktot}
\end{equation}
In the previous equation $m(a)$ is the mass of a grain of size $a$, $a_\mathrm{min/max}(r)$ are the minimum/maximum grain sizes at each radius and $n(a) \, \mathrm{d}a$ is the number of grains with sizes between $a$ and $a+\mathrm{d}a$. To reduce computational costs and better isolate the effects of the other variables in the analysis, we again adopt the commonly used power-law approximation for the grain size distribution: $n(a) \, \mathrm{d}a \propto a^{-q} \, \mathrm{d}a $ with $q = 3.5$, which is consistent with observations in protoplanetary disks \citep{2023Doi}. We note that the resulting size distribution in the \texttt{DustPy} simulations is, in any case, similar to a power-law with $q = 3.5$, so the effects of this assumption are small. With this choice of $q$, the value of $a_\mathrm{min}(r)$ has little effect on the opacity at mm-wavelengths and we fix it to $10^{-5}$ cm at all radii. At each radius, $a_\mathrm{max}(r)$ is defined as the maximum grain size that has non-negligible dust density (see Subsection \ref{ssec:dustpy}).

%observed a power law index of the grain size distribution ($q$) equal to 3.5 in the interstellar medium. In protoplanetary disks, grain growth and accretion processes might alter this distribution both globally and locally. Testing different values of $q$ is beyond the scope of this work and does not affect 

%Hence, we analyze the evolution of the power-law index in our \texttt{DustPy} simulations at different radial positions (see Appendix \ref{app:p} for more details). We find that, at most radii and in the peaks, the grain size distribution deviates from the canonical power-law approximation only at the upper end of the grain size spectrum (close to $a_\mathrm{max}$; see Appendix \ref{app:p} for more details). Instead, as expected, the small particles population in the gaps is better represented by higher values of $q$. Therefore, we simplify our analysis by assuming $q = 3.5$ while evaluating equation \ref{eq:ktot}. With this choice of $q$, the value of $a_\mathrm{min}$ has little effect on the opacity at mm-wavelengths and we fix it to $10^{-5}$ cm at all radii. 

To evaluate the dust opacities $k_\nu^\mathrm{a}(a)$ and $k_\nu^\mathrm{s}(a)$, as well as the particle masses, we adopted the "DSHARP" dust composition: a compact mixture of water ice, troilite, refractory organics, and astronomical silicates \citep{Birnstiel2018}. We compute the opacities with the Python package \texttt{dsharp$\_$opac}. We will not discuss different dust opacities in this work but we emphasize how different dust materials, such as amorphous carbons/silicates results in very different opacities and dust properties, first of all dust masses (Zagaria et al., in prep.). 

Note that, despite the approximations required to derive them,  \citet{Zhu2019} showed that both the equations \ref{eq:inu} and \ref{eq:scat}  are in excellent agreement with more complex and time-consuming radiative transfers simulations. Thus, we follow this analytical approach without testing more computational expensive radiative transfer models. 

\subsection{Simulated observations}
\label{ssec:simobs}
Through equation \ref{eq:inu}, we evaluate the intensity profiles of the dust continuum emission of both the compact and extended disk model at the following wavelengths: 0.45 mm, 0.87 mm, 1.29 mm, 3.07 mm and 7.46 mm. We also produce their 2D emission map, after assuming axisimmetry. To simulate the effect of the finite resolution of real observations on our models, we then apply to the multi-frequency 2D emission maps the \textit{simobserve} task of \texttt{CASA}  \citep{2022CASA}.

Given an input sky model, \textit{simobserve} accounts both for discrete uv-sampling of the ALMA/VLA configurations and the thermal noise. Given the different requirements on sensitivity of our multi-wavelengths and multi-facility observations, we neglect the thermal noise in our images. By doing this, we assume that the emission at all wavelengths is detected with high quality and signal-to-noise and the flux calibration error is the dominant source of uncertainties. Further comments about the limitations introduced in the SED analysis by limited sensitivity can be found in Subsection \ref{ssec:frank}. 

To test how the limited resolution and the beam smearing affect the results of the SED characterization, we produce mock observations with three different resolutions: 0.05", the maximum resolution achievable by the VLA in Band Q; 0.1", the maximum resolution of ALMA Band 1; and 0.2", a reasonable and time-efficient resolution for ALMA Band 1 observations of disks. 

We choose to position our mock sources at the distance (140 pc) and at the coordinates of HL Tau. At this declination (around +18:13:57), sources can be observed with a contained beam elongation by both ALMA and VLA.
\begin{table}[h!]
\caption{ALMA/VLA configurations used to simulate the multi-wavelength observations.}
\centering
\begin{tabular}{cccc}
\hline
$\lambda$ [mm] & Res 0.05" & Res 0.2" & Res 0.1" \\ \hline
0.45  & ALMA C7+4& ALMA C4+1& ALMA C5+2\\
0.87 & ALMA C8+5& ALMA C5+2& ALMA C6+3\\ 
1.29  & ALMA C9+6& ALMA C6+3& ALMA C7+4 \\ 
3.07 & ALMA C10+7& ALMA C8+5& ALMA C8+5\\ 
7.46  & VLA A+B& ALMA C9+6& ALMA C10+7\\ \hline
\end{tabular}
\vspace{0.10cm}
\label{table:config}
\vspace{-0.2cm}
\end{table}

The ALMA and VLA antennas configurations used to simulate the observations at the different frequencies and with different resolution are listed in Table \ref{table:config}. At each wavelength, we observe with both an extended and compact configuration. We follow the ALMA technical handbook for the time multipliers between configurations. We use a time multiplier of 0.25 between VLA configuration B and A. Each extended configuration has been observed for 3 hours in transit. 

We image the simulated observations with the \texttt{CASA} task \textit{tclean}. We employ the \textit{mtmfs} deconvolver \citep{Rau2011}, assuming that the frequency dependency of the emission follows a Taylor expansion to first terms (i.e., nterms = 2). We use \textit{briggs} weighting and multiple scales at 0 (point-source), 1, 3, 5 times the beam size. We define the pixel size to be $\sim$ 1/7 of the beam size. Different values of the robust parameter, from -1.0 to 0.5, were explored to produce a clean beam as close as possible to the expected resolution. To minimize any uncertainty due to convolution in the image plane, we also make use of an \textit{uv-taper} during the cleaning procedure to circularize the beam and approach the expected resolution. Finally, we smooth the clean image to the chosen resolution (0.05" or 0.1" or 0.2") using the \textit{imsmooth} task of \texttt{CASA}, assuring that the emissions at different frequencies are all convolved to the same beam.

We obtain azimuthally averaged radial intensity profiles of the mock observations by averaging the emission in concentric annuli (here rings since we assume face-on disks). To minimize correlated information in adjacent annuli, we define the radial width of each annulus as 1/3 of the beam size. The uncertainty of the radial profiles in each annulus is computed as the standard error of the mean: the standard deviation within the annulus ($\sigma_\mathrm{Annulus}(r)$) divided by the square root of the number of beams $N_\mathrm{Beams}$. In other words:
\begin{equation}
\overline{\sigma_\mathrm{Annulus}}(r) = \frac{\sigma_\mathrm{Annulus} (r)}{\sqrt{N_\mathrm{Beams}}} = \frac{\sigma_\mathrm{Annulus} (r)}{\sqrt{A_\mathrm{Annulus}/A_\mathrm{Beams}}},
\end{equation}
where $A_\mathrm{Annulus}$ and $A_\mathrm{Beams}$ are the area of the annulus and the beam respectively.

\section{SED analysis}
\label{sec:analysis}
We now aim to treat our synthetic multi-wavelength observations as real data. Specifically, we extract information about the dust content from our simulations by analyzing their spatially resolved SEDs. We then compare these measured values with the initial prescriptions used in our models, in order to assess the reliability of this procedure usually applied to real data.
%To measure the dust content of the compact and extended disk models and compare them with our prescriptions, we analyze the disk SEDs through a Bayesian approach. 
%At each radius, the mock intensity profiles at various wavelengths are matched with the physical model in equation \ref{eq:inu}, a function of the dust temperature $T_\mathrm{dust}$, surface density $\Sigma_\mathrm{dust}$ and maximum grain size $a_\mathrm{max}$. We, of course, make the same assumptions on dust composition and optical properties as we did in Subsection \ref{ssec:radiative}. 

Through a Bayesian approach, we compare, at each radius, the mock intensity profiles at various wavelengths to the physical model in equation \ref{eq:inu}, which is a function of the dust temperature $T_\mathrm{dust}(r)$, surface density $\Sigma_\mathrm{dust}(r)$ and maximum grain size $a_\mathrm{max}(r)$. We make the same assumptions on dust composition and optical properties as we did in Subsection \ref{ssec:radiative}.

We employ the code \texttt{emcee} \citep{ForemanMackey2019}, that implements the affine-invariant Markov-Chain Monte Carlo (MCMC) Ensemble Sampler of \citet{Goodman2010}, to estimate the posterior distribution of the model parameters at each radius. We use a standard log-normal likelihood function:
\begin{equation}
\ln{p(I_\nu(r) | \theta) }= -\frac{1}{2} \, \sum_{\nu} \left( \left(\frac{I_\nu(r) - I_\nu^\mathrm{model}(r, \theta)}{\sigma_{\mathrm{tot}, \nu}(r)} \right)^2 + \ln{(2\pi\, \sigma^2_{\mathrm{tot}, \nu}(r) )} \right),
\end{equation}
where $I_\nu(r)$ is the azimuthally averaged intensity profile at radius $r$ and $I_\nu^\mathrm{model}(r, \theta)$ is the model intensity in equation \ref{eq:inu} evaluated for the vector of model parameters $\theta = [T_\mathrm{dust}, \Sigma_\mathrm{dust}, a_\mathrm{max}]$.

In this work, we fix the power-law index of the grain size distribution $q$, introduced in equation \ref{eq:ktot}, to 3.5. This is consistent with how we simulated the multi-wavelength dust thermal emission (see Subsection \ref{ssec:radiative}). Since this parameter is strongly correlated with $a_\mathrm{max}$, we still perform a simple test of the ability of the SED analysis to retrieve both parameters (see Appendix \ref{app:p}).

The total uncertainty $\sigma_{\mathrm{tot}, \nu}(r) )$ is defined as:
\begin{equation}
\label{eq:uncert}
\sigma^2_{\mathrm{tot}, \nu}(r) ) = \sigma^2_\nu(r) + (\delta_\nu \, I_\nu(r) )^2,
\end{equation}
where $\sigma_\nu(r)$ is the standard error of the mean obtained while azimuthally averaging the intensity profiles (see Subsection \ref{ssec:simobs}). Instead $\delta_\nu$ is the flux calibration error. Following the ALMA Technical Handbook\footnote{\citet{ALMA2023}} and the Guide to Observing with the VLA\footnote{\citet{VLA2023}}, we set $\delta_\nu = 10\%$ for ALMA Band 9, 7, 6, 1 and VLA Q Band; instead, we adopt $\delta_\nu = 5\%$ for ALMA Band 3 data. 

While sampling, we allow the parameters to vary within uniform priors of ranges: $0 \leq T_\mathrm{dust}/\mathrm{K} \leq 100$,  $0 \leq (\Sigma_\mathrm{dust}/ \mathrm{g} \, \mathrm{cm}^{-2} ) \leq 30$ and $0 \leq (a_\mathrm{max}/ \mathrm{cm} ) \leq 100$. %Because of our choice of an uniform prior, we sample each parameter in linear - instead of logarithmic - scale to avoid biasing the results toward low values. 
We sample each parameter in linear - instead of logarithmic - scale. It is to be noted that, for the maximum grain size, this might not be the most appropriate choice. Grains larger than 10 cm do not emit efficiently in the millimeter fluxes we are analyzing. Therefore, a logarithmic sampling biased toward lower values or a narrower uniform prior are a better representation of our physical expectations. Here, we still choose a larger upper limit for the maximum grain size and a normal sampling to highlight when the parameters are unconstrained in our results.

As an example, Appendix \ref{app:sampling} shows that, if optically thin observations are included in an SED analysis, both a linear and a logarithmic sampling produce consistent results. Instead, if only marginally thick observations are considered, the complexity of the parameter space can cause the samplers to find a local convergence within a more confined prior or with a logarithmic sampling, leading to a solution that is reasonable but not necessarily accurate. To test the stability of the solution of an SED analysis, we suggest to double check the results with both a linear and logarithmic sampling or to explore the parameter space with other sampling techniques, such as a No-U-Turn Sampler (NUTS).

%since the complexity of the parameter space can cause the samplers to find a local convergence within more confined priors, leading to a solution that is reasonable but not necessarily accurate. 

Following \citet{macias2021characterizing}, we also impose a temperature prior based on the expected temperature profile of a passively irradiated flared disk in radiative equilibrium:
\begin{equation}
\label{eq:Tprior}
T_\mathrm{dust}(r) = \left( \frac{\psi \, L_\star}{8 \, \pi \, r^2 \, \sigma_\mathrm{SB}} \right)^{0.25}.
\end{equation}
In the previous equation $\sigma_\mathrm{SB}$ is the Stefan-Boltzmann constant and $\psi$ is the flaring angle that we vary uniformly between 0.01-0.06 (\citealt{huang2018disk}; \citealt{2001Dullemond}). We assume a normal distribution for the stellar luminosity $L_\star$ centered at 5 $L_\odot$ and with a conservative standard deviation of 50$\%$ $L_\star$. 

To assess convergence we follow \citet{Goodman2010} by introducing in our code criteria based on the \textit{integrated autocorrelation time}. After a burn-in of 1000 steps, we perform a maximum of 50 sampling, each with a total number of steps equal to 1000. At each sampling cycle, we evaluate the autocorrelation time. We define a chain converged if: i) they are longer than 100 times the estimated autocorrelation time, i.e. the walkers are independent; ii) for each parameter, the autocorrelation time at different cycles changes by less than 1$\%$, i.e. the convergence is stable. We choose to initialize 12 walkers (equal to 4 times the dimension of the problem), a larger number of walkers in this parameter space results in increasing autocorrelation times. We test different moves of the ensamble and we conclude that the default stretch move of \citet{Goodman2010} leads to faster convergence. 

The result of an \texttt{emcee} analysis are the marginalized probability distributions of the parameters. In the following, we will define as "results" of an SED analysis the value corresponding to the 50th percentiles of the distribution and, as errors, the difference between the 50th and the 16th and 84th percentiles: 50th$_{-(50{\mathrm{th}}-16\mathrm{th})}^{+(84\mathrm{th}-50\mathrm{th})}$. 

\section{Benchmark of the SED analysis}
\label{sec:benchmark}
\subsection{Different Band combinations}
\label{ssec:bands}
\begin{figure*}[h]{}
   \centering
   \includegraphics[width=1\textwidth]{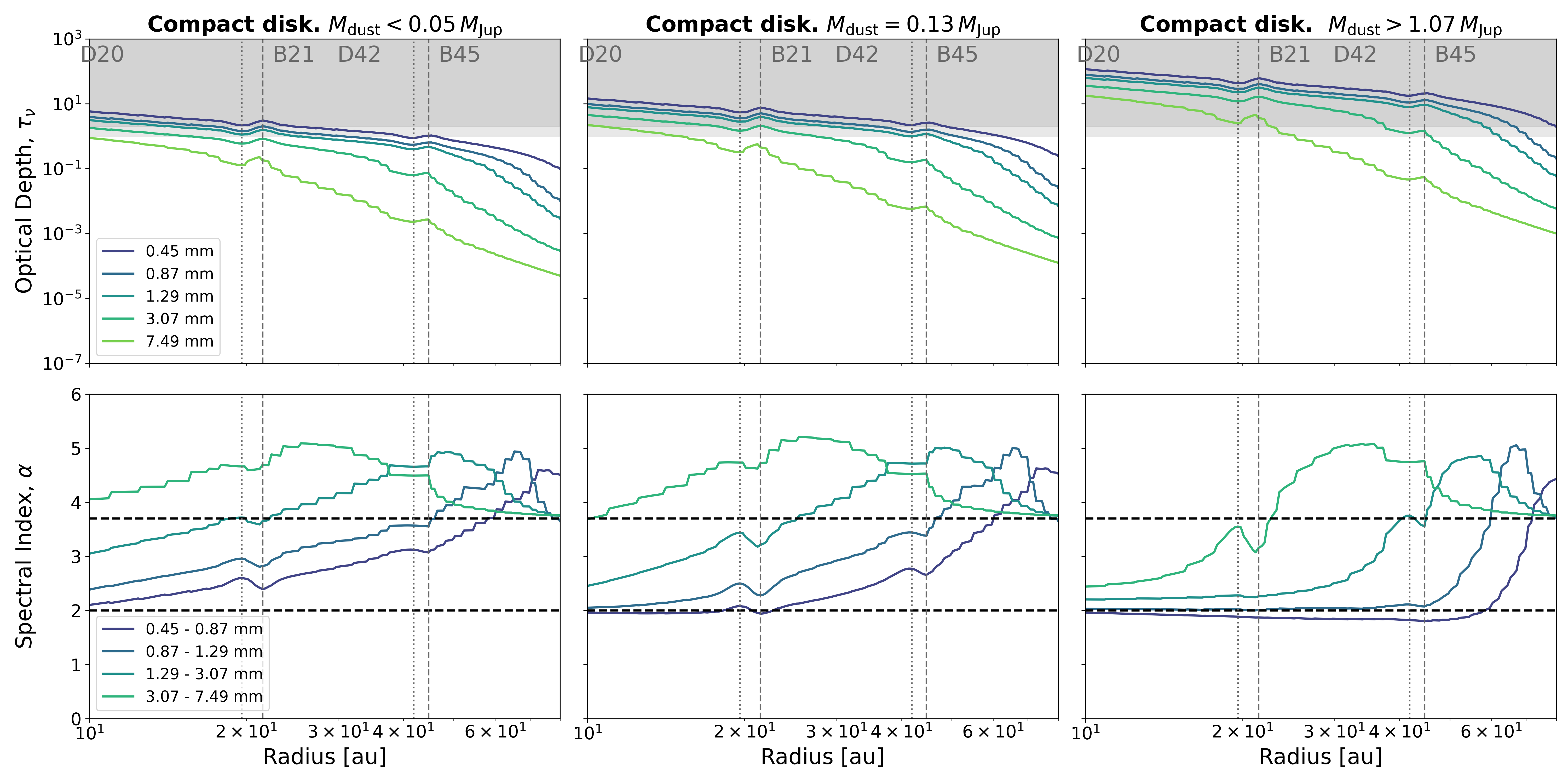}
   \includegraphics[width=1\textwidth]{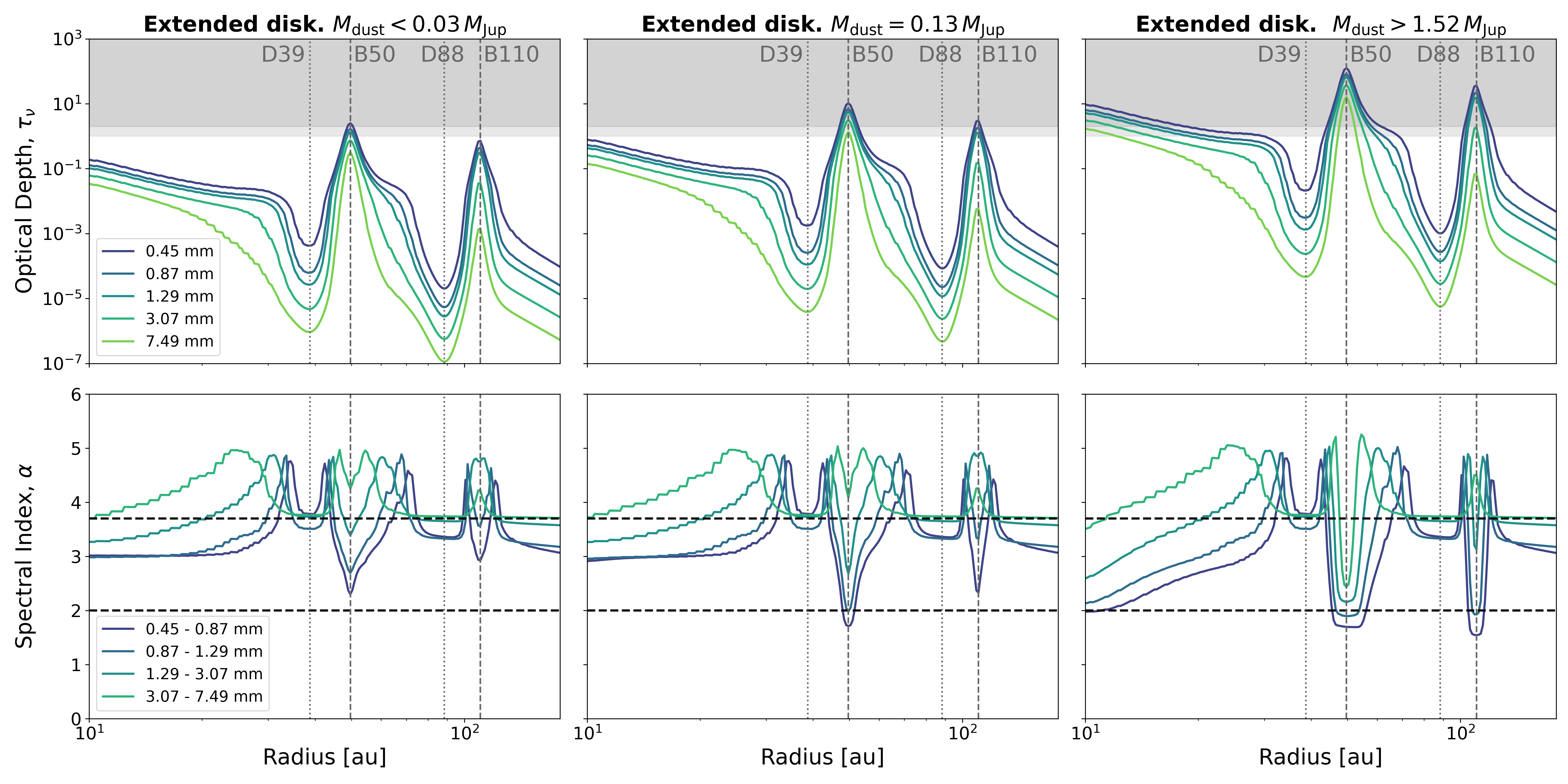}
   
   \caption{Optical depths (\textit{upper row}) and spectral indices (\textit{lower row}) computed for the compact (\textit{top figure}) and the extended disk (\textit{bottom figure}) for the three scenarios: i) dust-poor  disks with optically thin emission at both 3 mm and 7 mm (\textit{left}); ii) dust-average disks with optically thick emission at 3 mm and thin at 7 mm (\textit{middle}); and, iii) dust-rich disks with optically thick emission at both 3 and 7 mm (\textit{right}). The dust masses that give rise to these optical properties (for fixed maximum grain size and distance), are (from \textit{left} to \textit{right}): 0.06 $M_\mathrm{Jup}$, 0.14 $M_\mathrm{Jup}$ and 1.10 $M_\mathrm{Jup}$ for the compact disk; and, 0.03 $M_\mathrm{Jup}$, 0.13 $M_\mathrm{Jup}$ and 1.53 $M_\mathrm{Jup}$ for the extended disk. The vertical gray lines represent the position of the bright (B) and dark (D) rings in the dust surface density profile. The horizontal lines in the spectral indices plots indicate $\alpha$ = 3.7 (typical value in the ISM) and $\alpha$ = 2.0 (optically thick emission in the Rayleigh–Jeans limit). The dark gray shadow in the optical depth plots indicate $\tau > 2$, the light gray one  $\tau > 1$.}
    \label{fig:spectral_indeces}%
\end{figure*}
In the wide range of frequencies spanned by ALMA and VLA, the dust thermal continuum emission of a protoplanetary disk is characterized by very different optical depths that aid the reconstruction of the dust properties from a disk SED differently. In a simplified case where the dust opacity is dominated only by absorption and scattering is negligible (at mm/cm wavelengths this assumption is valid only for small particles $a_\mathrm{max} \leq 10^{-3} $ cm and might not be appropriate for protoplanetary disks; \citealt{birnstiel2011evolutiongasdustprotoplanetary}), equation \ref{eq:inu} simplifies to $I_\nu = B_\nu(T_\mathrm{dust}) \, (1-\mathrm{e}^{(-\tau_{\nu}^{a} / \mu)})$, where $\tau_{\nu}^{a}$ is the optical depth derived by the absorption opacity alone.

In this simplified scenario, optically thick emissions help get an accurate constraint on $T_\mathrm{dust}$. Indeed, if $\tau_{\nu}^{a} \gg 1$, the previous equation reduces to $I^\mathrm{thick}_\nu = B_\nu(T_\mathrm{dust})$ and we are able to measure the dust temperature free from degeneracies with the other dust properties. Instead, optically thin observations are sensitive also to the dust mass and particle sizes. If $\tau_{\nu}^{a} \ll 1$, $I^\mathrm{thin}_\nu \simeq B_\nu(T_\mathrm{dust}) \, \tau_{\nu}^{a} \simeq B_\nu(T_\mathrm{dust}) \, \Sigma_\mathrm{dust} \, k_{\nu}^{a}$.

At millimeter wavelengths, the spectral behavior of the absorption coefficient is usually parameterized as $k_\nu \propto \nu^\beta$. Therefore, in the Rayleigh–Jeans approximation, valid for disks at low optically thin frequencies, the optically thin intensity scales as $I^\mathrm{thin}_\nu \propto \nu^{2+\beta}$, and the spectral index of the emission between two millimeter wavelengths, $\alpha = \log{(I_{\nu 1}/I_{\nu 2})} / \log{({\nu_{1}}/{\nu_{2}})}$ spans from 2 (optically thick) to 2+$\beta$ (optically thin). Thus, in the thin case, the maximum grain size can be easily inferred from the measure of $\beta$ as $\beta = \alpha - 2$ (\citealt{Testi2014}; \citealt{Ricci2010}).

How the non-negligible dust scattering affects the previous picture is not trivial, since the intensity of an optically thick disk includes a dependence from the albedo if scattering is considered: $I^\mathrm{thick}_\nu \sim B_\nu(T_\mathrm{dust})\cdot \sqrt{1-\omega_\nu}$ \citep{Zhu2019}. Therefore, a complete SED analysis including absorption and scattering effects is important. 
\begin{figure*}
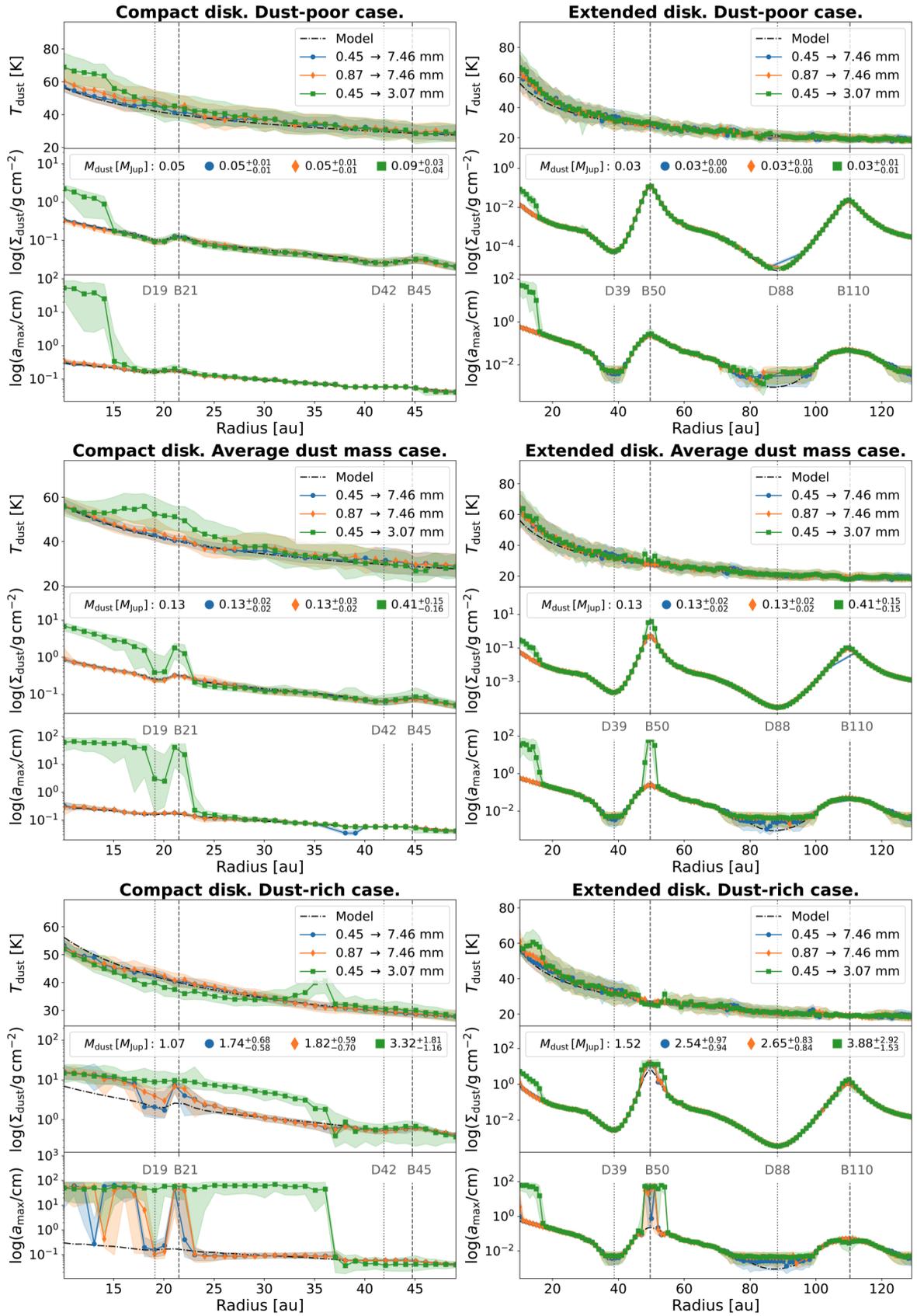
{}
   \centering
    \includegraphics[width=0.415\textwidth]{Final_High_Amax2median_posteriors.pdf}
    \includegraphics[width=0.415\textwidth]{Final_High_Amaxmedian_posteriors.pdf}
    
   \includegraphics[width=0.415\textwidth]
   {Final_Int_Amax2median_posteriors.pdf}
    \includegraphics[width=0.415\textwidth]{Final_Int_Amaxmedian_posteriors.pdf}
    
    \includegraphics[width=0.415\textwidth]{Final_Low_Amax2median_posteriors.pdf}
    \includegraphics[width=0.415\textwidth]{Final_Low_Amaxmedian_posteriors.pdf}
   
   \caption{Results of the three SED analyses for the compact (\textit{left}) and the extended (\textit{right}) disks. The three rows represent, in order: the dust-rich case (i), the average dust mass case (ii), and the dust-poor case (see Subsection \ref{ssec:bands}). Each subplot displays the dust temperature (\textit{top}), dust surface density (\textit{middle}) and maximum grain size (\textit{bottom}) measured from an SED analysis that includes noiseless, infinite resolution emissions from: 0.45 mm $\to$ 7.46 mm (in \textit{blue}); 0.87 mm $\to$ 7.46 mm (in \textit{orange}); and, 3.07 mm $\to$ 7.46 mm (in \textit{green}). The continuum line represents the 50th percentiles of the marginalized probability distribution of the parameters, the shaded area extends from the 16th to the 84th percentiles. The \textit{black} line indicates the dust properties of the models. %The vertical gray lines represent the position of the bright (B) and darks (D) gaps in the dust surface density profile.
   }
    \label{fig:bands}%
\end{figure*}

Even in a single spectral window the spectral properties of a disk dust emission are not uniform. The most recent observations of protoplanetary disks with ALMA at high angular resolution have revealed that dust substructures are ubiquitous, and the emission in the central parts of the disks, as well as in some of the bright rings, could be very optically thick even at long wavelengths (e.g., \citealt{Jin2016}; \citealt{Pinte2016}; \citealt{Liu2017}). In other words, for certain disk morphologies and dust masses, the 3 mm dust continuum emission could still be marginally thick and insufficient to the measurement of the dust properties from the SED. In these cases, longer-wavelengths but, unfortunately, lower resolution observations are to be included in the analysis (such as the 7 mm emission from ALMA Band 1 or VLA Band Q observations).

Here, we assess, for the two different disk morphologies we are simulating (compact, with shallow rings; and extended, double-ringed disk models), which ranges of dust masses result in marginally thick emission at 3 mm and 7 mm and how these optical depths affect the measurement of dust properties from the disk's SED. Note that we are assuming a reasonable distance of 140 pc (the distance of nearby star-forming regions such as Taurus and Ophiuchus), and a typical particle size distribution (mm-sized in the inner disk and rings and sub-mm in the outer disk)

To vary the optical depths of our simulated observations, we keep the maximum grain size profiles $a_\mathrm{max}(r)$, measured from our \texttt{DustPy} simulations (see Subsection \ref{ssec:dustpy}), fixed while manually rescaling the dust surface densities $\Sigma_\mathrm{dust}(r)$. We reproduce three scenarios: i) the case where the disks have enough dust mass that their emission in the inner disk (< 15 au)/brighter ring is marginally thick at both 3 and 7 mm (optical depths $\tau_{3, 7 \mathrm{ mm}} > 1-2$); ii) the more expected case of less massive but still detectable disks where the inner disk/brighter rings are marginally thick at 3 mm but thin at longer wavelengths ($\tau_{3 \mathrm{ mm}} > 1-2$, $\tau_{7 \mathrm{ mm}} < 1$); and, iii) a last case where the disks emission is thin at both 3 mm and 7 mm even in the inner disk and in the most dense substructures ($\tau_{3, 7 \mathrm{ mm}} < 1$).

For both the compact disk (\textit{top figure}) and the extended one (\textit{bottom figure}), the optical depths of the multi-wavelengths observations in the three scenarios are shown in the \textit{top rows} of Figure \ref{fig:spectral_indeces}. The \textit{right panel} show the dust-rich disk case (i), obtained by increasing the dust mass until the inner disk (for the compact disk) or the brighter ring (for the extended disk) have marginally thick 7 mm emission. Instead, the \textit{central} and \textit{left} panels, represent, respectively, the average dust mass case (ii) and the dust-poor case (iii). 

For both disk morphologies, the dust mass ranges that result in the three scenarios are almost the same. This is expected since the simulated disks have the same age, distance, initial mass and have a similar grain size profile. Since we choose typical values for all these properties, similar ranges are to be expected for the majority of protoplanetary disks.
The ranges of dust masses that lead to marginally thick 3 mm and thin 7 mm emissions are: $0.05 < M_\mathrm{dust} / M_\mathrm{Jup} < 1.07$ for the compact disk and $0.03 < M_\mathrm{dust} / M_\mathrm{Jup} < 1.52$ for the extended one. With the current ALMA and VLA sensitivities a disk that falls in the dust-poor picture (with masses smaller than 0.03/0.05 $M_\mathrm{Jup}$) is typically too faint to have detected 7 mm continuum emission at a distance of $\sim 140$ pc. On the other hand, recent dynamical mass measurements of the total mass of protoplanetary disks \citep{Martire_2024}, suggest that - for a standard gas-to-dust ratio of 100 - even the brightest disks do not typically exceed a dust mass of 1-2 $M_\mathrm{Jup}$. Consequently, we expect the majority of the detected protoplanetary disks to have dense, marginally thick 3 mm emission regions and, instead, optically thin 7 mm emission. 

Figure \ref{fig:spectral_indeces} also shows the spectral indices of the multi-wavelength intensity profiles computed in the three scenarios (\textit{bottom row}). To disentangle optical depth effects from the ones due to the limited resolution of the observations, here we compute the radial intensity profiles without applying \textit{simobserve}. Thus, these spectral indices have infinite resolution. Again, we note that the spectral indices obtained in the intermediate mass case (ii) are in agreement with the few real measurement available. A spectral index between Band 7 and 6 (0.87 mm and 1.29 mm) of $\sim$ 2 in the dense disk regions have been commonly observed in T-Tauri stars (see e.g., HL Tau, \citealt{carrasco2019radial}; CI Tau, Zagaria 2024 in prep.; and TW Hya, \citealt{macias2021characterizing}). A similar spectral index is expected between Band 7 and Band 9 (0.45 mm) since, at these frequencies, the emission is optically thick. \citet{sierra2021molecules} shows spectral indices between 100 GHz and 226 GHz (Band 3 to Band 6) between 2.5 and 3.5 within 50 au for the MAPS disks. Spectral indices between Band 4 (2.1 mm) and 8 mm (VLA Q+Ka Band) of 3.3 (at 50 au) and 3.5 (at 80 au) have been reported for HL Tau by \citet{carrasco2019radial}. 

To evaluate how the emissions at different optical depths aid the reconstruction of the dust properties, we perform, for each dust mass scenario, three SED analyses including, each time, a different combination of the intensity profiles at various wavelengths. In particular, we analyze the following Band combinations: a) from 0.45 mm to 7.46 mm; b) from 0.87 mm to 7.46 mm; and, c) from 0.87 mm to 3.07 mm. Note that we always include at least three wavelengths in the SED analysis to assure a good retrieval of the dust parameters. Degeneracies in the results of SED analyses have been observed by \citep{2024Li} when including only two wavelengths.

Case a) is a well-studied source for which we have very thick emission (Band 9 here but the same apply for ALMA Band 10/8 observations), intermediate emission (ALMA Band 7 and 6), marginally thin emission (ALMA Band 3 or Band 4) but also optically thin emission (ALMA Band 1 or VLA Q/Ka Band). Compared to this previous picture, in case b) we miss the very thick emission. Case c) is the more common case of a disk with available high-resolution Band 7, 6 and 3 observations. It is to be noted that while this band combination is more realistic (since it requires the least amount of wavelengths and does not include the expensive, longer ones), it is still uncommon to have these high-resolution multi-wavelengths observations: there are only 11 disks with available high-resolution (< 0.05") Band 7, 6, and 3 observations. 
 
The results of the three SED analyses for the compact (\textit{left}) and the extended (\textit{right}) disks are shown in Figure \ref{fig:bands}. The three rows represents, in order: the dust-rich case (i), the average dust mass case (ii), and the dust-poor case. Each subplot displays the dust temperature (\textit{top}), dust surface density (\textit{middle}) and maximum grain size (\textit{bottom}) measured from an SED analysis of: a) all the bands (in \textit{blue}, 0.45 mm $\to$ 7.46 mm); b) Bands 7, 6, 3 and 1 (in \textit{orange}, 0.87 mm $\to$ 7.46 mm); and, c) without the long wavelengths (in \textit{green}, 3.07 mm $\to$ 7.46 mm). The \textit{black} line indicates the dust properties of the models.

The conclusions of Figure \ref{fig:bands} are clear: in a dust-poor disk in which the 3 mm emission is already optically thin, longer-wavelengths observations are not needed to accurately reconstruct the dust properties from the disk SED. Indeed, both the broad (\textit{blue}) and narrow-band (\textit{green}) analysis in the \textit{last row} of Figure \ref{fig:bands} recover the expected dust temperature, surface density and maximum grain size almost everywhere. Only in the inner disk (< 15 au) our simulations have marginally thick 3 mm emission and we can distinguish the results obtained with and without including the 7 mm emission. Instead, for a protoplanetary disk with an average dust mass (\textit{second row}), the difference between including or not the 7 mm emission is non-negligible. The dust content in the dense regions of the disk, such as the inner disk and the rings, is robustly probed only including this optically thin emission. Lastly, in a dust-rich scenario (\textit{first row}), even lower frequencies than 40.0 GHz (7 mm) needs to be included in the SED analysis and none of the three band combinations we are testing is sufficient to accurately measure dust properties in the whole disk. This discussion is independent from the disk morphology for similar optical properties.

Figure \ref{fig:bands} also highlights that, by including a conservative but reliable prior (see equation \ref{eq:Tprior}), the dust temperature is always well constrained in an SED analysis independently from the optical depths of the observations. %On the other hand, the dust mass and the maximum grain size are constrained from the optical depth, a quantity that varies with frequency. \ev{%This also justifies why the B9 emission is not significantly improving the measurement of the dust temperature, even in the cases where the B7 and B6 observations are marginally thin (see Figure \ref{fig:spectral_indeces}). 
Contrary to the expectations of an absorption-only scenario (see the beginning of this Section), we found that completely optically thick observations do not aid in measuring the dust temperature. For a disk with an average dust mass, this suggests that Band 9 observations are not required. At millimeter wavelengths, the main contribution to opacity in B9 is scattering, making it less effective as a temperature tracer despite its high optical depth.

Note that, in all three dust mass scenarios, the maximum grain size in the first and second gaps of the extended disk morphology are not well constrained. This has a simple explanations: the dust population in these gaps has sizes of $\sim 10-100 \, \mu$m, one order of magnitude lower than the sub-mm wavelengths. Grains that are much smaller than the mean free path of the photons ($a < \lambda / 2\pi$) act like Rayleigh scatterers. Consequently, the opacities are insensitive to the grain sizes and we are not able to constrain them from an SED analysis (see Figure 4 of \citealt{Birnstiel2018}).

On top of including flux calibration errors while analyzing the multi-band SED, in Appendix \ref{app:flux}, we also investigate the potential impact of these errors by performing 50 separate SED analyses. In each analysis, the intensity profiles at the various wavelengths are rescaled within their respective flux calibration error margins. We observe that the spread in dust temperature measurements is less than 17$\%$. Additionally, the dust surface density and maximum grain size are nicely constrained.
Except near the very inner disk, the order of magnitude for both of these parameters is consistently matched.

\subsection{Different spatial resolutions}
\label{ssec:beams}
After testing how the different optical depths of the multi-wavelengths observations contribute to the measurement of dust properties from a disk SED, now we focus instead on the effect of the finite resolution of disks observations. As discussed in the previous Subsection, we expect the majority of protoplanetary disks to be characterized by marginally thick 3 mm emissions regions. Therefore, if we want to accurately measure the dust content even of the denser substructures, we need to include longer-wavelengths observations, such as the 7 mm emission, detected by ALMA in Band 1 or by VLA in Q Band. This comes with a limit: currently, ALMA Band 1 has a maximum resolution of 0.1", insufficient to resolve substructures in most disks (< 0.03" - 0.05"). High-resolution is partially reached at high frequencies by VLA (0.05" for Q Band, 0.06" for Ka) but with poorer image quality. 
\begin{figure}[h!]
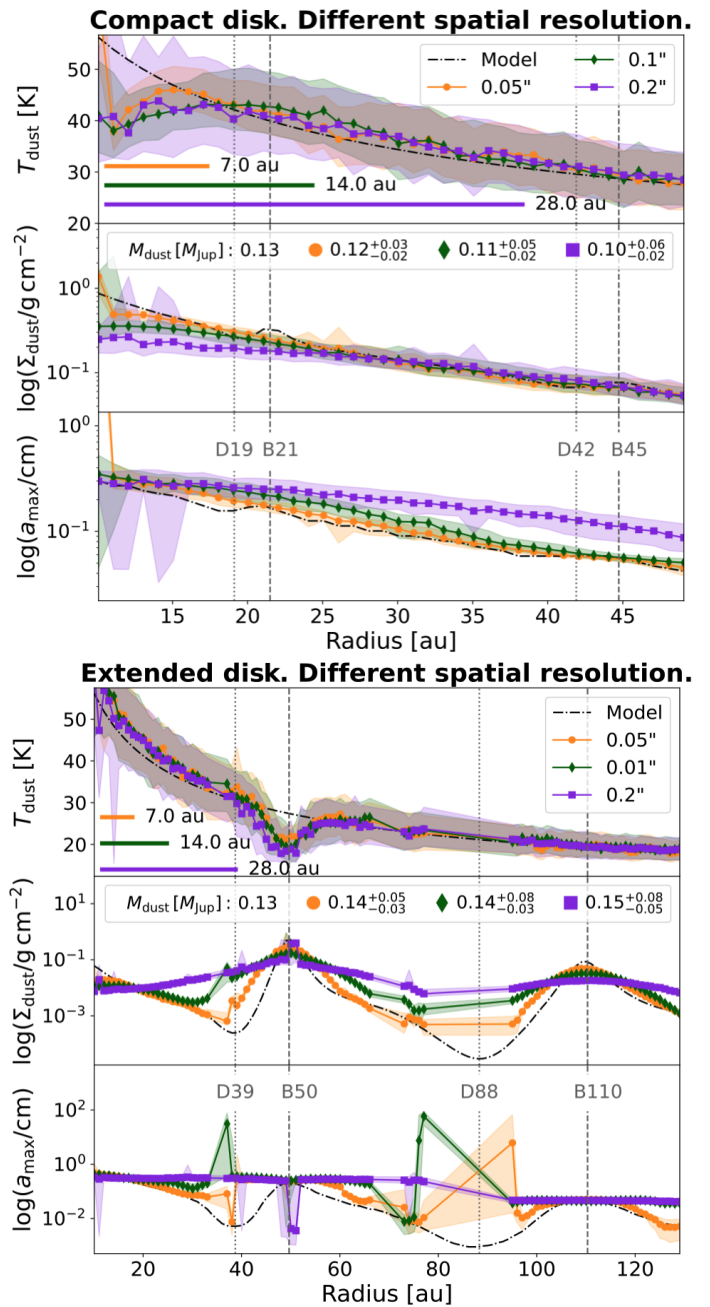

    \centering
    \includegraphics[width=\hsize]{Final_Beam_Amaxmedian_posteriors.pdf}
    \includegraphics[width=\hsize]{Final_Beam_Amax2median_posteriors.pdf}
    \caption{Dust temperature (\textit{top panel}), dust surface density (\textit{middle panel}) and maximum grain size (\textit{bottom panel}) measured from SED analyses of the 0.87 mm $\to$ 7.46 mm emissions for the average dust mass compact (\textit{top plot}) and extended (\textit{bottom plot}) disks. Each plot displays three solutions obtained from observations with different resolution: 0.05" (in \textit{orange}); 0.1" (in \textit{green}); and, 0.2" (in \textit{purple}). The continuum line represents the 50th percentiles of the marginalized probability distribution of the parameters, the shaded area extends from the 16th to the 84th percentiles. The \textit{black} line indicates the dust properties of the models. The vertical gray lines represent the position of the bright (B) and dark (D) gaps in the dust surface density profile. }
    \label{fig:beam}
\end{figure}

In this section, we compare the results of SED analyses of observations with different resolution. In particular, we focus on the following three resolutions: 0.05", the maximum resolution currently achievable by ALMA in Band 3 and by VLA in Q Band; 0.1", the maximum resolution achievable in ALMA Band 1; and, 0.2", a more realistic and time-efficient resolution for the current capability of ALMA. For a bright disk with a relatively low elevation from the ALMA site like HL Tau, like HL Tau, the maximum resolution achievable in ALMA Band 1 is 0.144". In order to detect the outer disk of HL Tau with a signal-to-noise greater than 5, a total observing time of 5.46 hours is required. By decreasing the resolution to 0.2", the total observing time is reduced by a factor of $\sim$4. Less bright disks like Elias 2-24, Tw Hya or PDS 70 can be all detected with a resolution of 0.2" and an rms of 5 $\mu$Jy in ALMA Band 1 in less than 6 hours. With this resolution, longer-wavelengths observations are in principle achievable for a sample of average brightness disks. 

Following the conclusions of the previous Subsection, here we analyze, for both the compact and the extended disks, at all three resolutions, the dust thermal emission from 0.87 mm to 7.46 mm of an average dust-massive disk (case ii). To simulate the effects of the observations on the multi-wavelength intensity profiles that we computed from equation \ref{eq:inu}, we apply the \textit{simobserve} task of \texttt{CASA} as explained in Subsection \ref{ssec:simobs}. The 7.46 mm emission at an angular resolution of 0.05" is simulated using a VLA antennas configuration, at lower resolution, we employ an ALMA Band 1 antennas configuration (see Table \ref{table:config}).

The dust temperature, surface density and maximum particle size measured from the SED analyses for the compact (\textit{top plot}) and extended (\textit{bottom plot}) disks are reported in Figure \ref{fig:beam}. Each plot displays the three solutions obtained from observations with resolutions: 0.05" (in \textit{orange}); 0.1" (in \textit{green}); and, 0.2" (in \textit{purple}). No solution is shown near the location of the second gap of the extended disk because the simulated observations of this deep substructure present negative intensity profiles. Thus, we exclude this region from the analysis.

Figure \ref{fig:beam} clearly highlights the need for high-resolution observations: for both the compact and extended disk, a medium-resolution SED analysis is insufficient to get an accurate measurement of the dust properties. The \textit{top plot} of Figure \ref{fig:beam} shows that the dust properties of the compact disk measured from 0.2" multi-wavelength observations generally underestimate those from the model in the inner disk and, at the same time, overestimate them in the outer parts. This is expected: at a distance of 140 pc, a resolution of 0.2" corresponds to 28 au, which means that the radial extent of the compact disk is resolved in only two beams. As a result of beam smearing, the flux of the inner disk is smeared toward the outer disk. Therefore, we have a reduction of the emission in the central parts while the emission of the outer disk is contaminated by the smeared flux. This leads to an underestimated dust temperature in the inner $\sim$ 20 au, that is otherwise consistent with expectation. The dust surface density is also only underestimated in the inner 20 au and robustly constrained elsewhere (with limited resolution); however, the maximum grain size has the opposite behavior: it is adequately constrained in the inner parts but severely overestimated outside. This results in underestimated optical depths in the inner disk and overestimated in the outer parts.

The beam smearing effect is reduced by increasing the resolution of the SED analysis. However, it is to be noted that even for the highest resolution (0.05", \textit{orange} line), we still see an underestimation of the dust temperature and surface density in the inner ~20 au. An even higher resolution would be required to resolve both the inner disk and the substructures in this compact disk with shallow gaps. The gap widths are $\sim$ 5 au for the first ring and  $\sim$ 7 au for the outer one and only marginally resolved even in the higher resolution case (0.05" = 7 au at 140 pc). 

The conclusions for the beam analysis comparison for the extended disk are different (see \textit{bottom plot} in Figure \ref{fig:beam}). This morphology presents a fainter inner disk and brighter dust rings than the compact disk and is almost three times more extended. Here, the beam smearing mainly affects the rings: because of the limited resolution the peak emission of the rings contaminates the flux coming from the gaps. This results, as evident from the \textit{purple} line (0.2") in the Figure, in uniform optical depths in the gaps and in the rings. In particular, we are not able to distinguish the dust surface density or the maximum grain size across the rings and gaps. A resolution of 0.1" is a great improvement to the SED characterization for this disk morphology (see \textit{green} line in Figure \ref{fig:beam}). While we still notice a non-negligible amount of flux contamination and a consequent overestimation of the dust surface density and maximum grain size in the gaps, here we witness different optical depths coming from the dark and bright regions of the disk. As previously discussed, this more ideal scenario is now achievable only for a few selected disks for which we have high-resolution long-wavelengths observations. With high-sensitivity ( $\gtrsim$ 50 hours) VLA Band Q observations, we can even push to a resolution of 0.05" (see \textit{orange} line in Figure \ref{fig:beam}) that, for a disk with a similar morphology and extent as the one we are simulating, provides accurate and robust measurements of the disk dust content. 

It is to be noted that, independently from the resolution, we are always able to estimate the maximum grain size at the peak of the substructures and the maximum level of grain growth in a disk. The same is true for the dust mass (see the labels in Figure \ref{fig:beam}): even though we miss spatial accuracy in the dust surface density, the total dust mass obtained by integrating the measured density profiles over the analyzed area remains accurate. This is true only if the optically thin observations are included in the analysis (see for a comparison Figure \ref{fig:bands}). A discussion on the uncertainties of dust mass measurements from optically thick observations is carried out in Section \ref{ssec:dustmass}.

\section{New approaches to the SED analysis}
\label{sec:improv}
In the previous Section, we commented on the necessity of including optically thin high-resolution observations in probing the dust content of protoplanetary disks. This limits the number of sources to which we can apply this dust characterization technique. While higher frequencies, higher resolution (< 0.05") observations are becoming more and more common for an increasingly wide sample of disks \citep{andrews2018disk}, high-quality ALMA observations still have a maximum resolution of 0.05" at 3 mm and 0.1" at 7 mm. With this resolution, we are unable to resolve the local dust accumulations in disks where the conditions for triggering streaming instabilities and forming planets are likely to occur.  

In this Section, we test on simulated observations techniques designed to handle a combination of lower and higher resolution intensity profiles and increase the spatial resolution of SED analyses. 

\subsection{The double-resolution analysis technique}
\label{ssec:double}
Here, we propose a new approach to the dust characterization method tailored for when a fraction of the intensity profiles of our multi-wavelength observations have significantly lower resolutions than the others. 

A requirement of SED analyses is to convolve all the observations to the lowest resolution. However, if some of the intensity profiles have medium to low resolution (> 0.05" - 0.1"), the results of the SED analysis are spatially unresolving substructures (see Figure \ref{fig:beam} and the discussion in Subsection \ref{ssec:beams}). To take advantage of the emission at all frequencies while resolving dust properties, we can use the unresolved results of this broad-band medium-resolution analysis, in particular the maximum grain size, as a prior of a second narrower-band analysis that includes in the SED only the highest resolution profiles.

\begin{figure}[h!]{}
   \centering
   \includegraphics[width=\hsize]{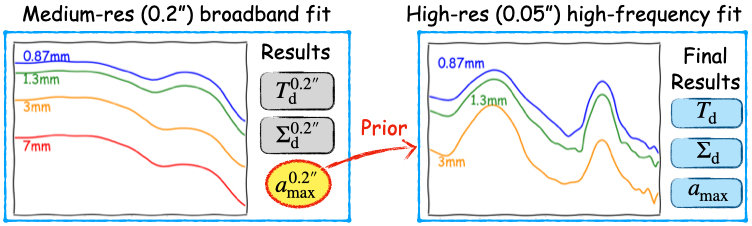}
   
   \caption{Sketch of the double SED analysis technique designed to handle a combination of lower and higher resolution multi-wavelengths observations of protoplanetary disks. }
    \label{fig:sketch}%
\end{figure}

Following the technical capabilities of ALMA, we test the realistic case of a disk with high-resolution (0.05") Band 7, 6 and 3 and medium-resolution (0.2") Band 1 observations. The double-resolution SED analysis technique for this scenario is sketched in Figure \ref{fig:sketch}: we perform a medium-resolution broadband SED analysis (0.87 mm $\to$ 7.46 mm with a resolution of 0.2") to measure the maximum grain size in the disk; with this information, we defined a prior on $a_\mathrm{max}(r)$ and implement it in a second analysis of the SED including only the 0.05" resolution 0.87 mm $\to$ 3.07 mm observations. We define the prior as an asymmetric Gaussian centered at the 50th percentiles of the marginalized probability distribution of $a_\mathrm{max}$ and with widths equals to two times the errors of the results (i.e., the differences between the 50th and 16th and the 84th and 50th percentiles). Defining the prior by interpolating the full density probability of the maximum grain size is equally valid but computationally expensive.  

This double-resolution approach can be applied even if we have more than one lower resolution observation, as long as the remaining high-resolution intensity profiles still have enough signal-to-noise for the results to not be affected by the decreased number of data in the SED. 
%If the lower resolution observations are (marginally) optically thick, the second SED analysis might benefit from a prior on the dust temperature as well. 

\begin{figure}[h!]
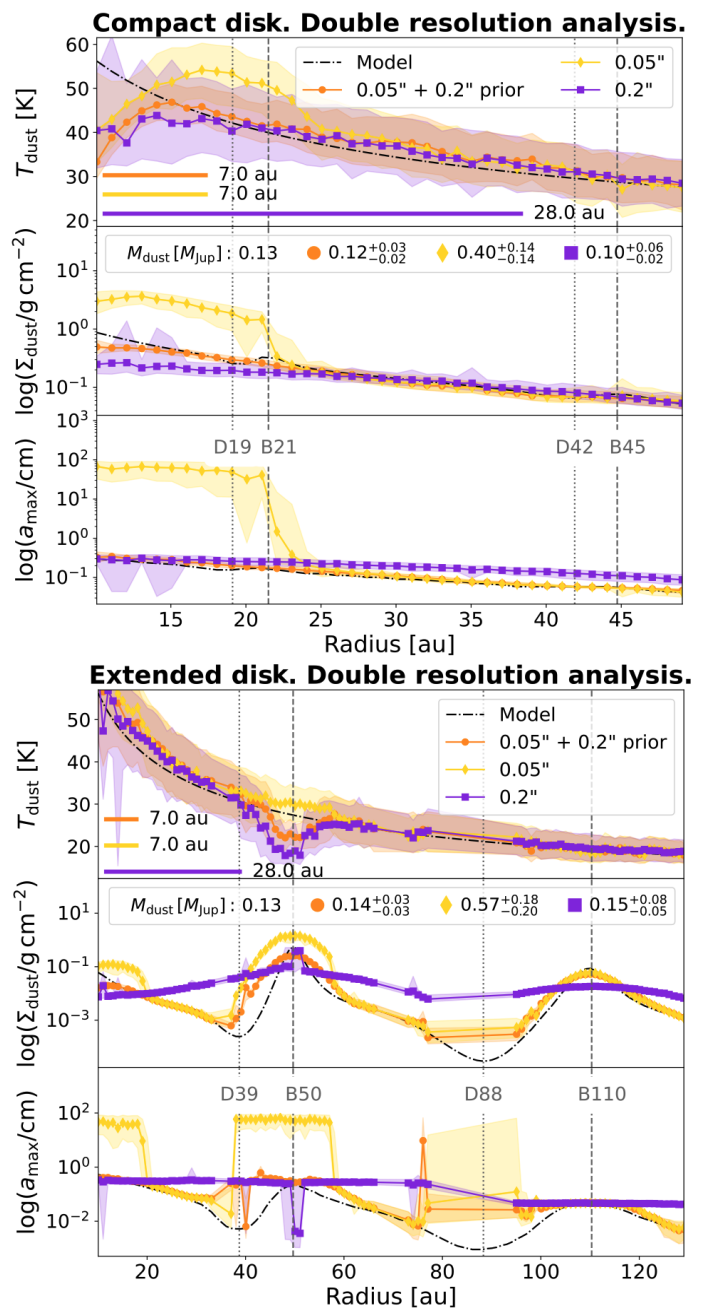

    \centering
    \includegraphics[width=\hsize]{Final_Double_Amaxmedian_posteriors.pdf}
    \includegraphics[width=\hsize]{Final_Double_Amax2median_posteriors.pdf}
    \caption{Dust temperature (\textit{top panel}), dust surface density (\textit{middle panel}) and maximum grain size (\textit{bottom panel}) measured from SED analyses for the compact (\textit{top plot}) and the extended (\textit{bottom plot}) disks. Each plot displays three solutions. In \textit{purple} we are showing the results of analyses from 0.87 mm $\to$ 7.46 mm with a resolution of 0.2"; in \textit{yellow} from 0.87 mm $\to$ 3.07 mm with a resolution of 0.05"; in \textit{orange} from 0.87 mm $\to$ 3.07 mm with a resolution of 0.05" but including a prior on the maximum grain size that is based on the results of the medium-resolution analyses (in \textit{purple}; see Subsection \ref{ssec:double}). The continuum line represents the 50th percentiles of the marginalized probability distribution of the parameters, the shaded area extends from the 16th to the 84th percentiles. The \textit{black} line indicates the dust properties of the models. The vertical gray lines represent the position of the bright (B) and dark (D) gaps in the dust
surface density profile.}
    \label{fig:double}
\end{figure}

The full potential of this double-resolution technique is shown in Figure \ref{fig:double}. Figure \ref{fig:double} shows in \textit{purple} the dust properties measured from the broadband lower resolution SED (0.87 mm $\to$ 7.46 mm with 0.2" of resolution) for both the compact (\textit{top plot}) and the extended disk (\textit{bottom plot}). As discussed in Subsection \ref{ssec:beams}, brightness peaks are smeared because of the limited resolution. This results in almost uniform dust surface density and maximum grain size over the disk extent (for the compact disk) or across gaps and rings (for the extended disk). Conversely, we succeed in radially resolving the dust content of both models if we only analyze high-resolution Band 7, 6 and 3 observations (\textit{yellow} solution). However, as we concluded in Subsection \ref{ssec:bands}, the dense regions of protoplanetary disks (i.e., the inner disks and the rings) have marginally thick emission even at Band 3, thus without optically thin information we are not able to probe their dust content. This changes by incorporating in this high-resolution analysis the optically thin information of the unresolved broadband SED (in \textit{purple}) through a prior. As depicted in \textit{orange} in Figure \ref{fig:double}, the optically thin information contained in the prior aids in measuring the dust content even in these denser regions without compromising resolution.

\subsection{Enhancing the resolution with \texttt{FRANK}}
\label{ssec:frank}
Alternatively, we can increase the spatial resolution of an SED analysis by individually enhancing the resolution of each radial intensity profile of the multi-wavelength observations. Unlike the double-resolution method, this approach does not necessarily require to reduce the number of profiles used in the SED analysis, allowing, in principle, for a better noise management.

To extract higher angular resolution intensity profiles from the multi-wavelength dust thermal emissions, we perform a non-parametric visibility modeling of the continuum visibilities using \texttt{Frankenstein} \citep{Jennings2020}. This tool, designed to recover axisymmetric disk structures at a sub-beam resolution, presents itself as an alternative to the traditional technique that we use to recover the brightness profile of our models from their interferometric observations: deprojecting and azimuthally averaging their \texttt{CLEAN} model images (see Subsection \ref{ssec:simobs}). During the cleaning procedure, a disk image is convolved with a beam that, as previously discussed and observed, smears and reduces the brightness of all the features of the source of size comparable to or smaller than the beam. In contrast, \texttt{Frank} infers the unconvolved brightness distribution by fitting the Real part of the visibilities as a function of the uv-distance. Because of the hypothesis of axisymmetry, the Imaginary part of the visibilities is not analyzed.

Again, we follow the current capabilities of ALMA by modeling the interferometric data of the 0.05" Band 7, 6 and 3 and 0.2" Band 1 observations. 

So far, we simulated noiseless observations. This allowed us to focus on the intrinsic contribution of each frequency to the SED analysis without the limitations introduced by the different noise levels of the various bands. Since the performance of a \texttt{Frank} fit is deeply influenced by the noise level, here we employ the \textit{add$\_$vis$\_$noise} task of \texttt{Frank} to simulate the effect of thermal noise on the visibilities. This function introduce a spread in the visibilities by adding Gaussian noise according to the data weights. After calibration, the weight of a couple of antennas ($i,j$) is defined as 1/$\sigma_{i,j}^2$, where $\sigma_{i,j}$ is the rms noise of a given visibility \footnote{\citep{nrao_casa_guide}}. If the antennas systemic temperature can be approximated to an average temperature for the interferometer, this noise can be related to the expected rms noise of the disk image ($\sigma_\mathrm{rms}$) as:

\begin{equation}
\sigma_{i,j} = \sigma_\mathrm{rms} \, \sqrt{\frac{N(N-1)}{2} \, \frac{t_\mathrm{source}}{t_{i,j}} \, n_\mathrm{pol}} = \sigma_\mathrm{rms} \, \sqrt{N_\mathrm{vis} \, n_\mathrm{pol}},
\label{eq:weights}
\end{equation}
where $N$ is the number of antennas, $N_\mathrm{vis}$ the total number of visibilities, $n_\mathrm{pol}$ the number of polarization (here = 1), $t_{i,j}$ the integration time, and $t_\mathrm{source}$ the time on source.

To manually introduce noise in the visibilities, we follow the weights definition of equation \ref{eq:weights}. We employ the \textit{ALMA Sensitivity Calculator}\footnote{\citep{alma_cal}} to evaluate the expected noise level of our observations for a source at HL Tau's declination after 3 hours of observing time. We adopted the same observing time while producing our noiseless simulations (see Subsection \ref{ssec:simobs}). 

We fit at each frequency the noisy visibilities in logarithmic brightness space, to reduce oscillatory artifacts and prevents negative values in the reconstructed brightness profile. We set the disk radius out to which the fit is performed ($R_\mathrm{max}$) equal to 1" for the compact disk and 2" for the extended disk, as recommended in \citet{Jennings2020}, this is greater than 1.5 times the disks outer edge. To ensure a sub-beam fit, we define the number of collocation points used in the fit ($N$) equal to 300. To minimize artifacts in the brightness profiles, we set $\alpha$ = 1.30 and $w_\mathrm{smooth} = 10^{-1}$. These two hyper-parameters mimic, respectively, the signal-to-noise (SNR) threshold below which the data are not used when fitting the visibilities, and the smoothness of the fit to the power-spectrum. \citet{Jennings2020} suggest varying $\alpha$ between 1.05 and 1.30 and $w_\mathrm{smooth}$ between $10^{-4}$ and $10^{-1}$, thus $\alpha$ = 1.30 ensures we are not including noise-dominated features in the fit, and $w_\mathrm{smooth} = 10^{-1}$ that we are not overfitting.  

To check whether we reached a sub-beam resolution, we compute the intrinsic angular resolutions of the \texttt{Frank} radial profiles. Following \citet{Sierra2024}, we generate synthetic visibility observations of a delta function centered at a radius of $r_0 = R_\mathrm{max}/2$ with the same uv-coverage of each data set. A non-zero center was chosen to avoid boundary effects. However, \citet{Sierra2024} report that negligible differences are found when using a different r0. We then fit the input profile with the same method and choice of hyper-parameters. We define the intrinsic resolution as the broadening, i.e. the full width half maximum (FWHM), of the brightness profile reconstructed by \texttt{Frank}.

\begin{figure}[h]{}
   \centering
   \includegraphics[width=\hsize]{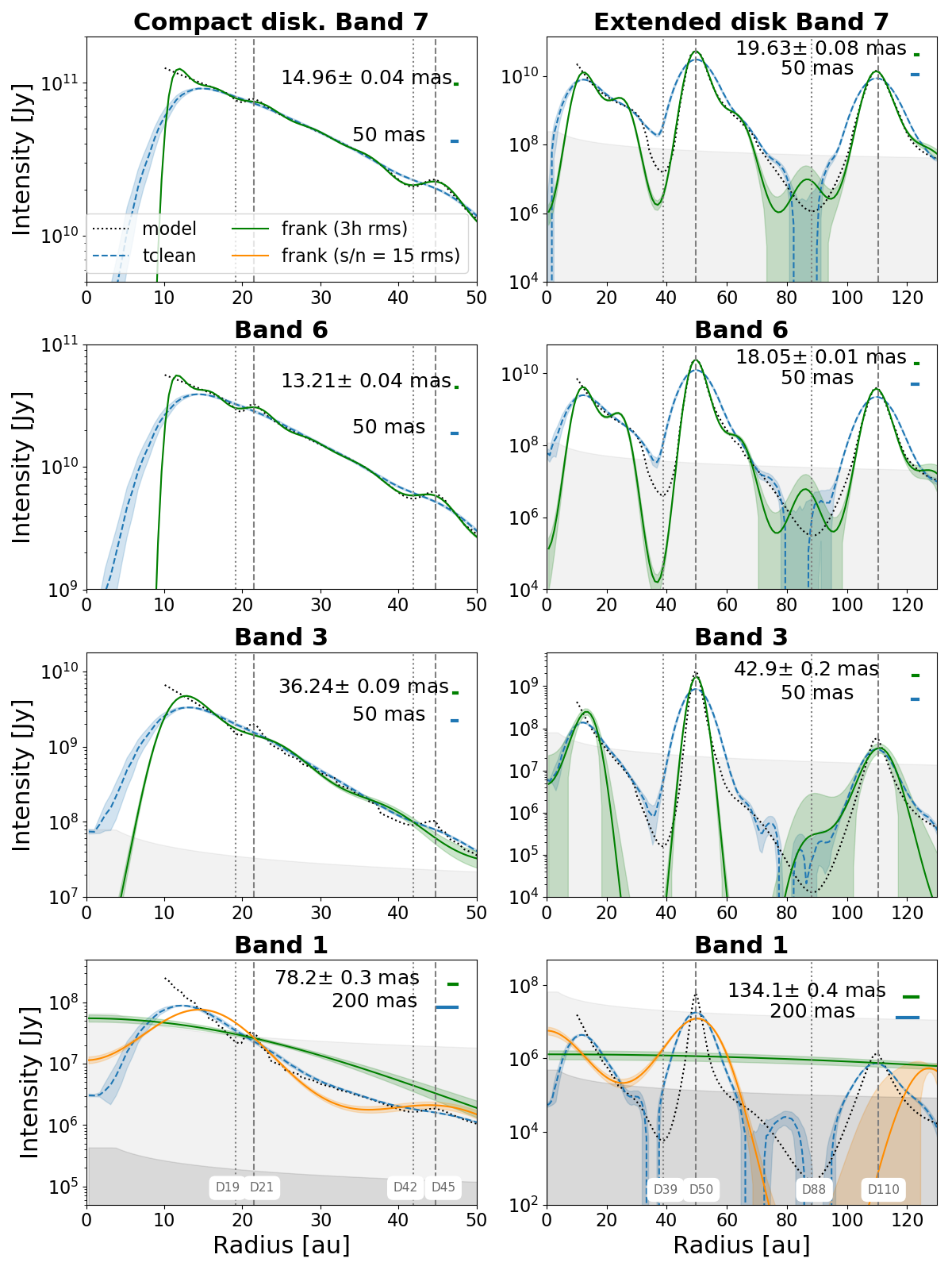}
   \caption{\texttt{Frank} fits of the multi-band observations of the compact (\textit{left}) and extended (\textit{right}) disk models. The fits to the visibilities with the expected noise level after 3 hours of observing time are shown in \textit{green}. The light gray shadow in each panel is the corresponding radially averaged noise level. The model intensity profiles are plotted in \textit{black} and the \textit{tclean} profiles in \textit{blue}. For Band 1 (\textit{bottom panel}), the fits to the visibilities with an azimuthally averaged signal-to-noise of 15 in the outer ring are also shown in \textit{orange} and the corresponding noise level is represented as a dark gray shadow. The vertical gray lines in each panel represent the position of the bright (B) and dark (D) gaps in the dust surface density profile. In each panel the intrinsic resolution of the \texttt{Frank} fits are reported.}
    \label{fig:frank}%
\end{figure}

The \texttt{Frank} fits of the multi-band observations of the compact (\textit{left}) and extended (\textit{right}) disk models with the expected noise level after 3 hours of observing time are shown in \textit{green} in Figure \ref{fig:frank}. The light gray shadow in each panel represents the corresponding radially averaged noise level. In the same Figure, the model intensity profiles are plotted in \textit{black} and the \textit{tclean} profiles, i.e., the profiles obtained by averaging the clean simulated observations (see Subsection \ref{ssec:simobs}), in \textit{blue}. 

Various considerations follow from Figure \ref{fig:frank}. The sensitivity requirements to reach a satisfying signal-to-noise are extremely different in the various ALMA bands. With three hours of observing time, the outer ring of the compact disk is detected with a radially averaged signal-to-noise of $\sim$ 330 in Band 7, $\sim$ 180 in Band 6, $\lesssim$ 5 in Band 3, and non detected in Band 1. It is to be noted that, even for non-detection, \texttt{Frank} still produce a result that, for an unresolved compact disk, could be confused with a physical one (see \textit{bottom left} panel of Figure \ref{fig:frank}). Similar SNRs are observed for the extended disk morphology. 

Following these non-detections, we visually inspected the minimum radial signal-to-noise in the outer ring of both the compact and extended disk required to produce \texttt{Frank} profiles from the Band 1 observations that approximate the model. For both morphologies, signal-to-noise ratios lower than 15 are insufficient. With the current ALMA capabilities, these low noise levels in Band 1 observations requires times on source of respectively 35 h for the compact disk and 27 h for the extended one. Such expensive observations are still not enough for an accurate reconstruction of the 7.46 mm intensity profile (see the \textit{orange} solutions in the \textit{bottom panels} of Figure \ref{fig:frank}). The stricter requirement in the SNR with respect to the Band 3 profiles is likely caused by the lower uv-coverage of the Band 1 observations. Note that we are increasing the signal-to-noise of the observations by increasing the weights, but the number of uv-points are fixed to the one produced from a 3 h observation. This is equivalent to binning the visibilities, an action that is typically adopted in faint observations to increase the SNR (see \citealt{Sierra2024}).

Artifacts in the \texttt{Frank} profiles are seen even for the higher resolution Band 3, 6 and 7 observations, independently from the signal-to-noise and despite the conservative values of the $\alpha$ hyper-parameter. The bright Band 7 and 6 observations are characterized by a bump in the intensity profile in the inner disk of both the extended and compact morphologies, suggesting that this is probably the high resolution response of \texttt{Frank} to the abrupt inner cut-off of the disk model at $\sim$ 10 au. Less expected bumps are found at the center of the compact disk in the Band 3 emission and in the first ring of the extended disk in Band 7 and 6. Additionally, we observed that different choices of hyper-parameters result in different profiles.

In each panel of Figure \ref{fig:frank}, we are also reporting the intrinsic resolution of the \texttt{Frank} profiles. We tested the robustness of these values against changes of the $N$ and $R_\mathrm{max}$ parameter. Since here the disks geometry is well known, the uncertainties on these values are low. The improvement in resolution compared to the tclean profile of the Band 7 and 6 observations is significant: $\gtrsim 3$ for the compact disk and $\gtrsim 2.5$ for the extended one. This is not the case for the fainter Band 3 observations (a factor $\sim$ 1.4 improvement for the compact disk, and $\sim$ 1.2 for the extended one). With high signal-to-noise ratio but limited uv-coverage, the \texttt{Frank} Band 1 profiles have an intrinsic resolution of 0.0782" ($\sim$ 2.5) and 0.1341" ($\sim$ 1.5) for the compact and extended disks. 

\begin{figure}[h!]
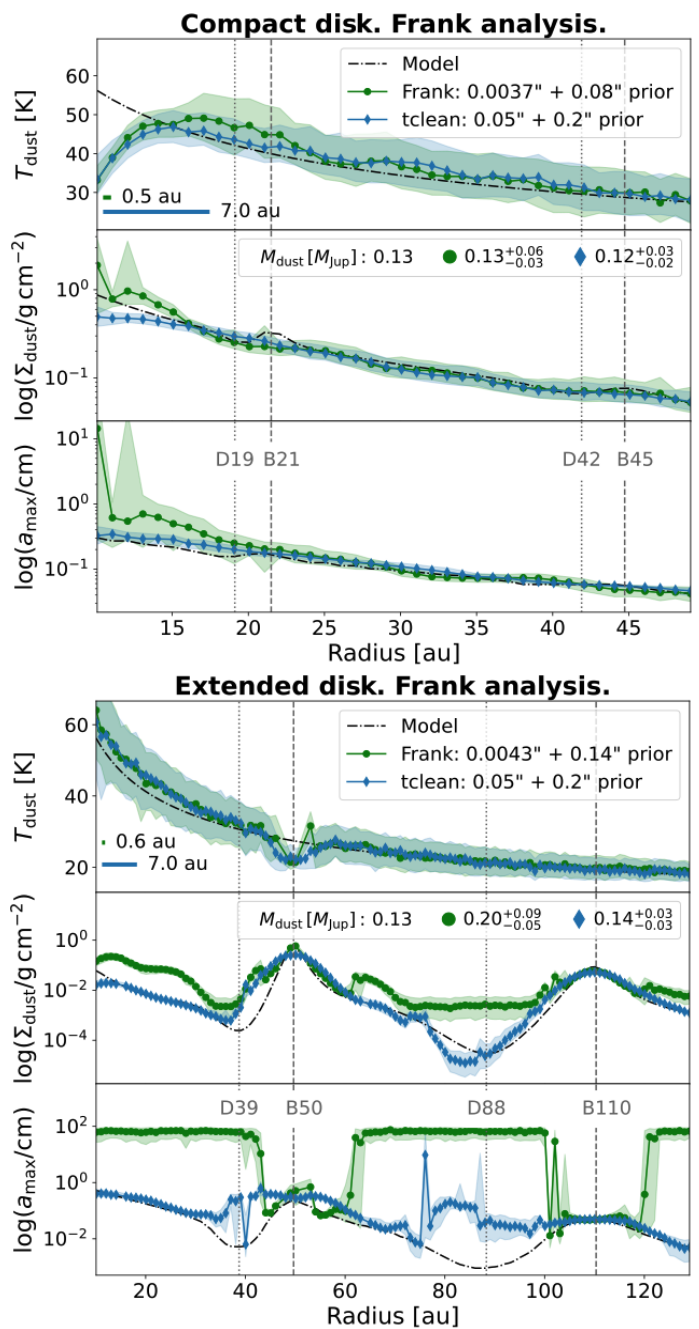
{}
   \centering
   \includegraphics[width=\hsize]{Final_Frank_Amaxmedian_posteriors.pdf}
   \includegraphics[width=\hsize]{Final_Frank_Amax2median_posteriors.pdf}
   \caption{Dust temperature (\textit{top panel}), dust surface density (\textit{middle panel}) and maximum grain size (\textit{bottom panel}) measured from SED analyses for the compact (\textit{top plot}) and the extended (\textit{bottom plot}) disks. Each plot displays the results of the double-resolution technique presented in Subsection \ref{ssec:double} applied on: the \texttt{Frank} intensity profiles (in \textit{green}), and the \textit{tclean} intensity profiles (in \textit{blue}). The continuum line represents the 50th percentiles of the marginalized probability distribution of the parameters, the shaded area extends from the 16th to the 84th percentiles. The \textit{black} line indicates the dust properties of the models. The vertical gray lines represent the position of the bright (B) and dark (D) gaps in the dust surface density profile.}
    \label{fig:frank2}%
\end{figure}

Since the increase in resolution of the \texttt{Frank} fits of the optically thin observations is still insufficient to resolve substructures, we apply the double-resolution technique on the \texttt{Frank} intensity profiles with Band 1 again as the lowest resolution profile (see Subsection \ref{ssec:double}). In this scenario, we are still reducing the number of data in the SED. The results are shown in \textit{green} in Figure \ref{fig:frank2}. For a comparison, we also display the results of the double-resolution technique applied on the \textit{tclean} profiles (\textit{blue}). 

For the compact disk, the difference between analyzing the \texttt{Frank} and the \textit{tclean} profiles is negligible. The improvement in resolution of the higher frequencies profiles (0.037" instead of 0.05", the intrinsic resolution of the Band 3 \texttt{Frank} fit) is not significant in resolving the substructures of this small disk. Unexpectedly, the important artifacts introduced in the Band 1 \texttt{Frank} fit are not affecting the dust properties determination. This might suggest that the Band 1 flux - instead of the spatially resolved observation - is sufficient in helping reconstructing the dust properties in the higher resolution SED analysis. This is also supported by the conclusions of the previous Subsection (see Figure \ref{fig:double}): lower-resolution Band 1 observations can still aid the dust content reconstruction when used as a prior of a narrowband higher resolution analysis. 

Similar results are found for the more complex morphology of the extended disk model. Despite the artifacts of the \texttt{Frank} profiles, the Band 1 information included in the prior are aiding the reconstruction of the dust properties in the non-noise-dominated rings. In the noise dominated regions of this disk, the optical depth is not constrained and no information on the maximum grain size or on the dust density can be gathered.

\section{Discussion}
\subsection{Measuring dust masses from millimeter emissions}
\label{ssec:dustmass}
More than 1000 protoplanetary disks in several nearby star-forming-regions have been surveyed with ALMA in the 211 GHz–373 GHz frequency range, where the dust emission and the CO isotopologues are bright (e.g., \citealt{pascucci2016steeper}; \citealt{ansdell2016alma}; \citealt{barenfeld2016alma}). These measured millimeter fluxes have been converted into dust masses via \citep{Hildebrand}:
\begin{equation}
\label{eq:mass}
M_\mathrm{dust} = \frac{1}{\mu}\frac{F_\nu \, D^2}{\overline{k_\nu} \, B_\nu(\overline{T}_\mathrm{dust})},
\end{equation}
where $F_\nu$ is the dust flux at frequency $\nu$, $B_\nu(\overline{T}_\mathrm{dust})$ is the Black-Body emission at the dust temperature $\overline{T}_\mathrm{dust}$ and frequency $\nu$, $\mu$ is the cosine of the disk inclination which has been assumed equal to 1 for most of the published surveys, $D$ is the distance of the source and $\overline{k}_\nu$ is the dust absorption opacity. The overline indicates that the dust properties are averaged over the disk extent.

Equation \ref{eq:mass} can be easily derived from the thermal emission of a 1D razor-thin vertically isothermal slab (equation \ref{eq:inu}) under the assumption of optically thin emission, but spectral indices between Band 7 and 6 of $\sim$ 2, consistent with optically thick emission, have been commonly observed in T-Tauri disks (e.g., \citealt{carrasco2019radial}; \citealt{macias2021characterizing}).

%Equation \ref{eq:mass} can be easily derived from the thermal emission of a 1D razor-thin vertically isothermal slab (equation \ref{eq:inu}) under two main assumptions: i) the scattering at frequency $\nu$ can be neglected; and ii) the emission at frequency $\nu$ is optically thin. Neither of them are verified for the Band 7/6 dust emission of protoplanetary disks. The albedo at 0.87 mm of a compact grain composed of a mixture of water, troilites, crystalline organics and astrophysical silicate ("DSHARP" dust composition) with a size of 1 mm is 0.9. Spectral indices between Band 7 and 6 of $\sim$ 2, consistent with optically thick emission, have been commonly observed in T-Tauri disks (e.g., \citealt{carrasco2019radial}; \citealt{macias2021characterizing}).

Through self-consistent radiative transfer models at 0.88 mm and 1.33 mm, \citet{liu2022underestimation} observed that this traditional dust mass measurement method can underestimate the mass by a factor of a few to several hundreds, depending on the optical depth along the line of sight set mainly by the true dust mass, disk size, and inclination. Here, we complement \citet{liu2022underestimation} results with an analysis of dust mass estimates obtained via the multi-wavelength analysis of simulated disks with different optical depths and within different frequency ranges.

\begin{figure}[h!]{}
   \centering
   \includegraphics[width=\hsize]{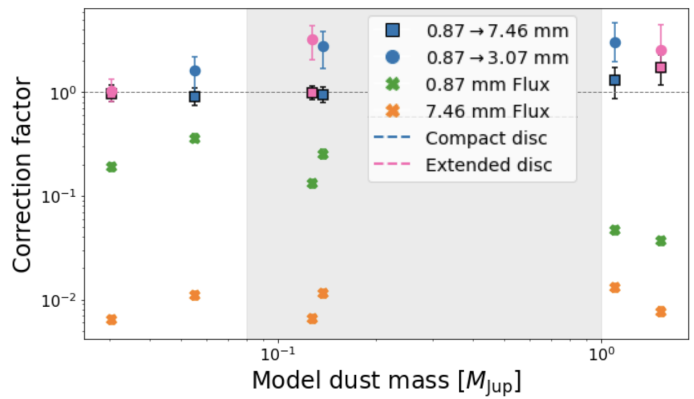}
   \caption{Correction factor, i.e. the ratio between the measured and the model dust mass, for various measurement techniques. \textit{Circles} refers to the masses measured from SED analyses including the emission from 0.87 mm to 3.07 mm, in \textit{blue} for the compact disk model and in \textit{pink} for the extended disk morphology (see Section \ref{sec:data}). With a \textit{black-bordered square} we are showing the masses measured from SED analyses that also include the 7.46 mm emission. The correction factors that we obtain by measuring the dust mass from the 0.87 mm (\textit{green cross}) and 7.46 mm (\textit{orange cross}) fluxes through equation \ref{eq:mass} are also displayed. The gray shadow highlights the range of dust masses that we can typically expect from ALMA detections of T-Tauri disks (scenario ii; see Subsection \ref{ssec:bands}). The horizontal dashed line marks a accurately estimated dust mass.}
    \label{fig:mass}%
\end{figure}

In Figure \ref{fig:mass}, we show the correction factor, i.e. the ratio between the measured and the model dust mass, that we obtain from our SED analyses. To investigate this technique ability to accurately measure dust masses for different optical depths, we again refer to the three scenarios we introduced in Subsection \ref{ssec:bands}:  i) a dust-rich ($M_\mathrm{dust} \gtrsim 1.6 M_\mathrm{Jup}$), ii) a dust-average ($0.06 M_\mathrm{Jup} \lesssim M_\mathrm{dust} \lesssim 1.6 M_\mathrm{Jup}$), and iii) a dust-poor ($M_\mathrm{dust} \lesssim 0.06 M_\mathrm{Jup}$) disk model. The results obtained from SED analyses of the emission from 0.87 mm to 3.07 mm are displayed with a \textit{circle} in \textit{blue} for the compact disk model and in \textit{pink} for the extended disk morphology. With a \textit{black-bordered square} we are showing the results of SED analyses that also include the 7.46 mm emission. Note that the dust masses have been measured from SED analyses of multi-band observations with infinite resolution. As discussed in Subsection \ref{ssec:beams}, the total dust mass is always recovered in SED analyses even if the intensity profiles are unresolved. The gray shadow highlights the range of dust masses that we can typically expect for ALMA observations of T-Tauri disks (scenario ii; see Subsection \ref{ssec:bands}). The horizontal dashed line marks an accurately estimated dust mass.

For a comparison, we are also showing in Figure \ref{fig:mass} the correction factors that we obtain by measuring the dust masses from the 0.87 mm (\textit{green cross}) and 7.46 mm (\textit{orange cross}) fluxes through equation \ref{eq:mass}. We follow \citet{2013Andrews} in defining a stellar-luminosity based average dust temperature: $\overline{T}_\mathrm{dust} = 25 \, (L_\star/L_\odot)^{0.25}$ K. In the literature, the dust average opacities are typically computed as the absorption opacity of a grain of 1 mm size (\citealt{2013Andrews}; \citealt{1990Beckwith}). For the "DSHARP" composition and assuming a power-law index of the grain size distribution $q  = 3.5$ (see Subsection \ref{ssec:radiative}), this equals 3.5 cm$^{2}$/g at 0.87 mm and 0.1 cm$^{2}$/g for 7.46 mm. We vary the flux within typical flux calibration errors (10$\%$ for both 0.87 mm and 7.46 mm) to estimate the uncertainties of these masses.

%From Figure \ref{fig:mass}, it is evident that classical flux-based mass measurements are not accurate. For a dust-rich disk, the masses derived from both the 0.87 mm and 7.46 mm fluxes underestimate the model for both compact and extended disk morphologies. This is expected, as at these high optical depths, even the 7.46 mm emission remains optically thick, concealing a significant amount of dust. For disks with average dust masses (scenario ii), the 0.87 mm emission is at least marginally optically thick, but the long-wavelength 7.46 mm flux is optically thin. This is not evident from Figure \ref{fig:mass}: for dust masses of $\sim$ 0.13 $M_\mathrm{Jup}$, the mass derived from the 7.46 mm flux is still underestimated by a factor $>2$. Similarly, for dust-poor disks where both the 0.87 mm and 7.46 mm fluxes are optically thin, flux-based mass measurements remain underestimated by $\sim 1$ order of magnitude at 0.87 mm and $\sim 2$ at 7.46 mm. Altogether, these findings highlight that the uncertainties in this mass measurement technique are being severely underestimated.

From Figure \ref{fig:mass}, it is evident that, independently from the optical depth of the observations and the disk morphology, classical flux-based mass measurements are not accurate. This is expected for the 0.87 mm flux (\textit{green cross}). The 0.87 mm emission remains marginally optically thick even for a dust-poor disk, concealing a significant amount of dust and leading to a severely underestimated dust mass. On the other hand, we expect the optically thin 7.46 mm flux to provide a better constraint of the dust content. This is not observed in Figure \ref{fig:mass}. The masses derived from the 7.46 mm flux (\textit{orange cross}) are even more underestimated than the ones derived from Band 7. This highlights another important simplification of this flux-based mass measurement technique: the assumption of a single arbitrary dust opacity instead of a more consistent value modeled at each radius. By assuming a grain size of 1 mm (close to the model's maximum grain size of $\sim 3$ mm), we are severely overestimating the dust opacity in Band 1, which in turn results in an underestimated dust mass. Indeed, according to Mie theory of dust opacity, the 7.46 mm flux presents a peak in the dust opacity at a grain size of the order of $\lambda/2\pi \sim 1$ mm \citep{Birnstiel2018}. 

Very different dust masses are measured from SED analyses. If optically thin observations are not included in the SEDs, the dust mass is overestimated - instead of being underestimated as in the case of flux-based masses - and, more importantly, not robustly constrained. This is the case for the dust masses derived for the dust rich disk models with any combination of wavelengths, as well the 0.87 mm $\to$ 3.07 mm analysis for the dust-average scenario. This behavior can be understood by looking at Figure \ref{fig:bands}, which shows the probability distribution of dust parameters obtained from SED analysis for both compact and extended disks across all three dust mass scenarios. When only optically thick observations are considered, the SED becomes insensitive to dust opacities, leading to different combinations of $\Sigma_\mathrm{dust}$ and $a_\mathrm{max}$ having similar probabilities. This implies a broad marginalized distribution for $\Sigma_\mathrm{dust}$, and thus an unconstrained dust mass. The overestimation of the masses is a consequence of linear sampling and the limited prior assumptions.  

On the contrary, optically thin observations in the SED analysis results in dust masses that are both accurate and robustly constrained. This is the case for the masses derived from the 0.87 mm $\to$ 7.46 mm (\textit{square}) SED analyses of the dust-average and dust-poor disk models as well as the 0.87 mm $\to$ 7.46 mm (\textit{circle}) SED analysis of the extended dust-poor disk (\textit{pink}). For the compact disk (\textit{blue}) in the dust-poor scenario, the dust mass obtained from the 0.87 mm $\to$ 3.07 mm SED analysis is slightly overestimated. As shown in Figure \ref{fig:bands}, the dense inner disk of this compact morphology remains marginally optically thick even at low dust masses.

To conclude, the results presented in Figure \ref{fig:mass} demonstrate that the SED analysis is a significantly more effective method for measuring dust masses in protoplanetary disks, provided that optically thin observations are included. While this analysis is more expensive than simply measuring dust masses from millimeter fluxes, the fact that the spatial resolution of these optically thin observations does not critically affect the derived dust mass estimates (see Subsection \ref{ssec:beams}) makes the method more feasible for samples of disks, as high-resolution observations are not strictly required. 
At the same time, it is also evident that an analysis that relies only on optically thin fluxes at a single wavelength is insufficient (see \textit{orange crosses} in Figure \ref{fig:mass}) since this method relies on assumptions about an average dust temperature and, in particular, an average dust opacity. %Due to the varying morphologies, ages, chemistry, and levels of grain growth in protoplanetary disks, dust opacities can differ significantly. 
%Therefore, a more complex and tailored approach to averaging opacities is needed, along with a more accurate method for estimating uncertainties.

\subsection{Extending SED analyses with the ALMA WSU and the ngVLA}
\label{ssec:multi-wave}
In most of the published SED analyses of protoplanetary disks, the longest wavelengths incorporated are high- to medium-resolution (0.05"-0.1") marginally thick observations at 2 mm - 3.6 mm (ALMA Band 3/4). Alongside the more frequently observed Band 6/7, fully optically thick data from ALMA Bands 9/10 are occasionally included.

The benchmarking of the dust characterization method on simulated observations, as shown in Section \ref{sec:benchmark}, indicates that fully optically thick observations do not significantly enhance the measurement of dust properties if an accurate prior can be included. This is in contrast with what was observed by \citep{2019Kim} while performing a synthetic ALMA multi band analysis to find the best band configuration for constraining the dust content of TW Hya. The authors found that the dust properties of the disk were better constrained when including ALMA Band 10/9 observations. However, it is to be noted that their radiative transfer model does not account for scattering. As discussed in Section \ref{ssec:bands}, at millimeter wavelengths the main contribution to opacity in B9 is scattering. Consequently, in the presence of scattering, the emission in Band 9 could be not as good as a temperature tracer as suggested by its high optical depth. Nevertheless, we note that including Band 9/10 observations in a dust characterization analysis might aid noise management by providing an additional point in the SED (see Subsection \ref{ssec:bands}).

More importantly, our analysis in Subsections \ref{ssec:bands} and \ref{ssec:beams} leads to a clear conclusion: to accurately constrain the dust content of protoplanetary disks through SED analyses, high-resolution optically thin observations must be included. For the typical T-Tauri disks we detect with ALMA, ($0.07 M_\mathrm{Jup} \lesssim M_\mathrm{dust} \lesssim 1.4 M_\mathrm{Jup}$), and under the assumption of a "DSHARP" dust composition (see Subsection \ref{ssec:radiative}), this requires wavelengths longer than $\sim$ 3 mm and resolutions $\lesssim$ 0.05". The wavelength at which the emission transitions from optically thin to optically thick can be approximately identified from the spectral indices. As discussed at the beginning of Subsection \ref{ssec:bands}, the spectral index of the emission at millimeter wavelengths spans from 2 for optically thick emission, to 2 + $\beta$ for optically thin emission, with $\beta$ being the power index of the absorption coefficient. This picture is complicated in a non-trivial way by dust scattering. However, spectral indices above a $\sim 3$ threshold, which are sensitive to substructures, are expected for completely optically thin observations (see Figure \ref{fig:spectral_indeces}).

As previously discussed, the lack of longer wavelength observations of protoplanetary disks in the SED analyses can be traced back to the fact that dust thermal emission at these wavelengths can only be observed with high-resolution by the VLA. The limited sensitivity and image quality of this facility constrains the analysis to bright disks visible at VLA declinations. In fact, SED analyses incorporating high-resolution VLA data are currently available for only two protoplanetary disks:  HD 163296 (\citealt{guidi2022distribution}) and HL Tau (\citealt{carrasco2019radial}). 

The new capabilities of ALMA Band 1 to produce high-quality and time-efficient long-wavelengths observations of average-brightness sources open a new window in dust characterization of protoplanetary disks. Although the maximum resolution of ALMA Band 1 (0.1") remains insufficient for resolving substructures, the double-resolution analysis we are proposing in Subsection \ref{ssec:double} effectively integrates optically thin information from these medium-resolution observations into a higher resolution framework, allowing for high-resolution results. 

With the upcoming Wide Sensitivity Upgrade (WSU), ALMA will further enhance our ability of measuring dust properties even of the faintest protoplanetary disks. The initially double, and eventually quadruple bandwidth upgrade will boost the continuum sensitivity of ALMA's receivers by a factor of 3 up to 6 \citep{carpenter2020almadevelopmentprogramroadmap}. This is crucial for easy detection of faint optically thin emission. Moreover, the WSU comes with a new access to long wavelengths: ALMA Band 2. Band 2 is designed to have the widest frequency coverage of any interferometric (sub)millimeter facility, from 67 GHz to 116 GHz, encompassing the Band 3 frequency range and extending to a maximum wavelength of 4.48 mm. This new Band is key for characterizing dust of faint disks, allowing for marginally thick ($\sim$ 3 mm) observations even of the faintest sources. While it does not extend as far as Band 1 in terms of wavelengths, a maximum wavelength of 4.48 mm can still provide sufficiently optically thin observations for faint sources with the advantage of increased sensitivity and resolution. 

An ongoing study led by the European Southern Observatory (ESO) also addresses the feasibility of extending ALMA's maximum baselines to as long as 32 km \citep{carpenter2020almadevelopmentprogramroadmap}. The corresponding increase in resolution at higher frequencies (e.g., ALMA Bands 6 and 7) will complement the milliarcsecond resolution that the next generation VLA (ngVLA) will provide at longer wavelengths ($\sim$ 3–9 mm) \citep{wilner2024keysciencegoalsgeneration}. This synergy between the two facilities will be critical, as it will allow us to resolve and characterize even the more compact and fainter protoplanetary disks, which constitute the majority of the population, including systems similar to the one that formed our solar system.

The increase in resolution provided by these facilities will also allow us to probe the dust content of even the most inner regions of protoplanetary disks, where rocky planets supposedly form. However, we know that at wavelengths longer than 6 mm, emission from ionized gas or other non-dust mechanisms can contaminate the dust emission originating from the inner parts of the disk \citep{carrasco2019radial}. To remove this contribution, high-resolution observations over a wide range of long wavelengths are needed. The VLA and the next-generation Very Large Array (ngVLA) will be fundamental in providing these observations in the Northern Hemisphere, in the same way the Square Kilometre Array (SKA) will allow us to access the Southern Hemisphere. 

With this boost in resolution across a wide wavelength range, we will finally have the capability to probe the fine structures and dust distributions in these smaller, less luminous disks, enabling a deeper understanding of the early stages of planet formation in a broader range of environments. 
%%%%%%%%%%%%%%%%%%%%%%%%%%%%%%%%%%%%%
%Then I think we should talk about how ngVLA will be very good here, but that it will need to provide high res at 3 mm too, and that ALMA will need to also likely increase the offered resolution and sensitivity so that there's a good frequency coverage (i.e., having 1 mas resolution at a single wavelength will not allow us to characterize the dust at that resolution, we need this resolution at least at 3-7 mm, and possibly at even higher frequency). Also, even the two resolution method can still only work on relatively bright sources, so ngVLA will be needed if we want to do this on weaker and more compact sources (which are more representative of the whole population).

\section{Summary and conclusions}
Spatially resolved multi-wavelength studies of the dust thermal emission of protoplanetary disks are essential for understanding the initial stages of planet formation. By accurately studying both the particle size distribution and the dust surface density in these disks, we can identify where and when the conditions necessary to trigger streaming instabilities are met.

In this work, we benchmark the state-of-the-art dust characterization technique, based on SED analyses at different radial locations, on simulated multi-band data. We include completely optically thick observations (ALMA Band 9, 0.45 mm), commonly observed 0.87 mm and 1.29 mm data (ALMA Band 7 and 6), marginally thick 3 mm emissions (ALMA Band 3); and, a longer wavelength 7 mm intensity profile. We design two disks model, a compact disk with shallow gaps (similar to TW Hya), and an extended double-ring morphology (a proxy for HD 163296).

We first test which bands aid the reconstruction of the dust properties from the SED, varying the optical depths of the observations by manually rescaling the dust mass. We find that, wavelengths longer that 3 mm are needed to accurately reconstruct the dust properties in dense disk regions such as the inner disks and the rings. Surprisingly, fully optically thick observations do not significantly enhance the measurement of dust properties, but they might aid with noise by providing an additional point in the SED.
We verified this for disks simulated at a distance 140 pc, assuming a "DSHARP" dust composition, and for dust masses between $\sim$ 0.06 $M_\mathrm{Jup}$ and 1.6 $M_\mathrm{Jup}$. Higher dust masses are rare even for the brightest disks, masses lower that $\sim$ 0.1 $M_\mathrm{Jup}$ are not detected with the current sensitivities of the more advanced radio facilities, such as ALMA and VLA. 

As a second test, we compare the dust properties derived from SED analysis with different resolutions: 0.05", the maximum resolution achievable by ALMA in Band 3 and by VLA in Q Band ($\sim$ 7-8 mm); 0.1", the maximum resolution of ALMA Band
1 in Cycle 11; and, 0.2", a more realistic and time-efficient resolution for the current capabilities of ALMA. We observe
that, in order to resolve substructures and characterize the radial variation of dust properties in the rings and gaps, high-resolution ($\lesssim$ 0.05") observations are necessary. For coarser estimates of the total dust mass and maximum grain sizes at ring locations, lower resolution (0.2") SED analyses are sufficient. Thus, analysis of multi-band fluxes can still be relevant to study the mass budget and the level of grain growth of samples of protoplanetary disks; but only the advent of milliarcseconds resolution facilities such as the ngVLA will allow us to actually resolve substructures and probe the planets' birth environments even in the more compact and fainter disks, which constitute the majority of the population.  

Because of the limited sensitivity and image quality achievable at long-wavelengths (7-8 mm) by the VLA, we propose a new approach to dust characterization that effectively integrates optically thin information from medium-resolution (0.2") ALMA Band 1 observations into a higher resolution framework. We demonstrate that the information obtained from a medium-resolution SED analysis that includes long-wavelength observations, can be used as a prior of a second narrower band SED analysis of only higher frequency, higher resolution intensity profiles. This technique opens a new window in the dust characterization of faint disks that will be further expanded with the upcoming ALMA's Wide Sensitivity Upgrade. This upgrade will boost the sensitivity of ALMA’s receivers. Additionally, ALMA Band 2 will extend up to 4.85 mm, which can still provide sufficiently optically thin observations for faint sources with the advantage of increased
sensitivity and resolution.

We also test how the flux reconstruction tool \texttt{Frankenstein} can help increasing the resolution of an SED analysis by individually enhancing the resolution of each radial intensity. We note that, even for high signal-to-noise, \texttt{Frank} introduces artifacts in the fitted profiles. Thus, these analyses should be carefully treated.

Finally, we discuss on the total dust mass that we derive from the SED analyses and compare it with the traditional method of deriving dust masses from (sub-)millimeter fluxes. Accurate dust mass measurements can be retrieved from the SED only if optically thin tracers are included. If not, the dust masses are not robustly constrained and generally overestimated. On the contrary, dust masses measured from the 0.87 mm fluxes can be underestimated by one order of magnitude, since the high optical depths at these wavelengths can hide substantially high amount of dust. We also observe that the accuracy of the results is independent from the spatial resolution of the SED within the explored range (0.05"-0.2"). We observe that mass derived from the 7.46 mm fluxes are also not accurate. A single-wavelength flux-based mass measurement method still relies on assumptions about average dust temperatures and opacities. Due
to the varying morphologies, ages, and levels of grain growth
in protoplanetary disks, dust opacities can differ significantly. A more complex and tailored approach to averaging
opacities is needed, along with a more accurate method for estimating uncertainties.

\begin{acknowledgements}
We thank the anonymous referee for the helpful report.
\end{acknowledgements}

%-------------------------------------------------------------------

\bibliographystyle{aa}
\bibliography{bibliography}

\begin{appendix}
\section{Fitting $q$: the power index of the grain size distribution }
\label{app:p}

As discussed in Subsection \ref{ssec:radiative}, a commonly used power-law
approximation for the grain size distribution in protoplanetary disks is $n(a) \, \mathrm{d}a \propto a^{-q} \, \mathrm{d}a $.

\citet{Mathis1977} observed a power law index of the grain size distribution ($q$) equal to 3.5 in the interstellar medium. In protoplanetary disks, however, grain growth and accretion processes can modify this value, both globally and, more significantly, locally. Higher values of $q$ are often found in disk gaps, which are populated by small dust particles, while in the rings $q$ values closer to 3.5 are observed, consistent with fragmentation processes (e.g., \citealt{macias2021characterizing}).

In this work, we assume a uniform value of $q = 3.5$, as this approximation does not affect the conclusions of our benchmarking analysis. Nonetheless, because of the strong correlation between $a_\mathrm{max}$ and $q$, here we assess the ability of the SED analysis method to retrieve both parameters. We do this by comparing the results of two analyses: one with $q$ fixed at 3.5, and another with $q$ as a free parameter. 

\begin{figure}[h!]{}
   \centering
   \includegraphics[width=\hsize]{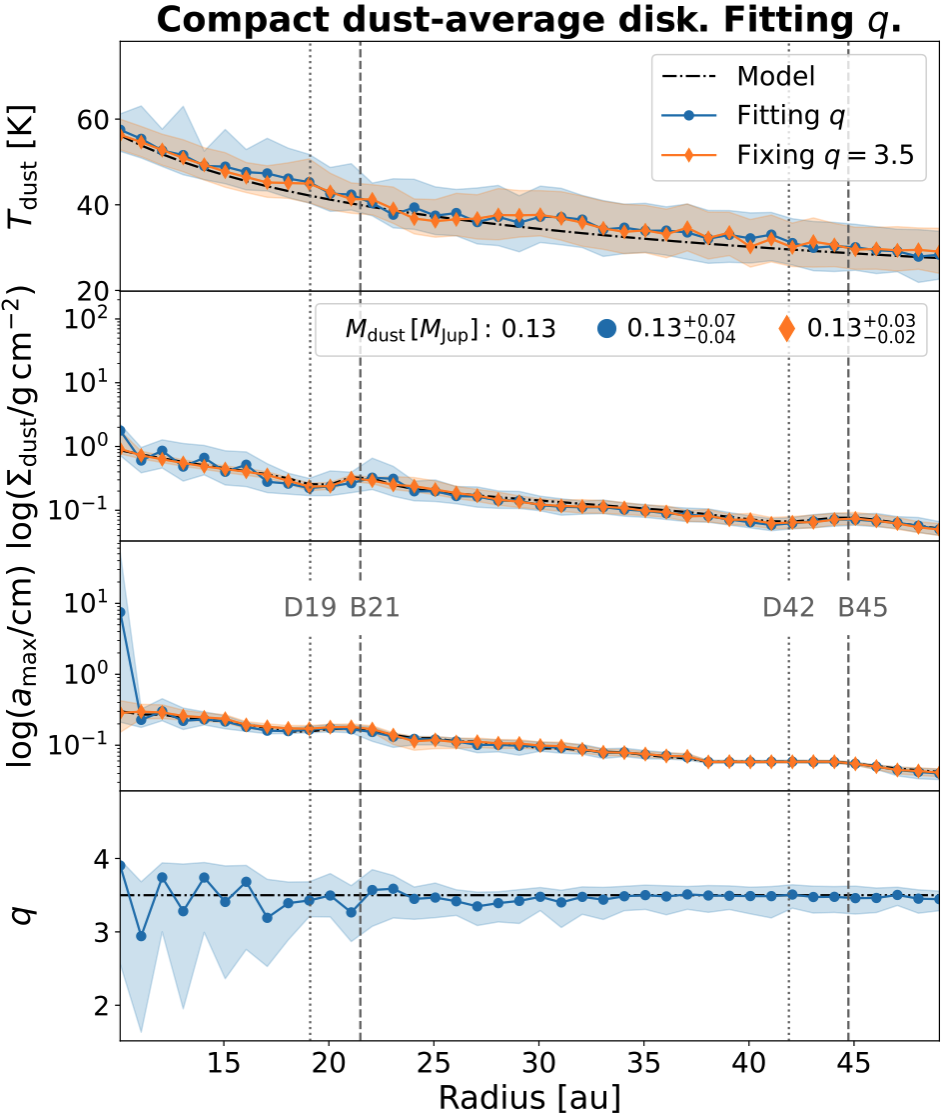}
   
   \caption{Dust temperature (\textit{first panel}), dust surface density (\textit{second panel}), maximum grain size (\textit{third panel}), and power index of the grain size distribution (\textit{fourth panel}) measured from an SED analysis of the compact disk model (see Subsection \ref{ssec:bands}). The SED includes emission from 0.87 mm to 7.46 mm, simulated with infinite resolution. The \textit{blue} solution is obtained considering the power index of the grain size distribution ($q$) as one of the parameter of the SED analysis. Instead, the \textit{orange} solution is the result of an SED analysis fixing $q$ = 3.5, the expected value in our simulations. The continuum line represents the 50th percentiles of the marginalized probability distribution of the parameters, the shaded area extends from the 16th to the 84th percentiles. The \textit{black} line indicates the dust properties of the models. The vertical gray lines represent the position of the bright (B) and dark (D) gaps in the dust surface density profile.}
    \label{fig:p}%
\end{figure}

\begin{figure}[h!]{}
   \centering
   \includegraphics[width=\hsize]{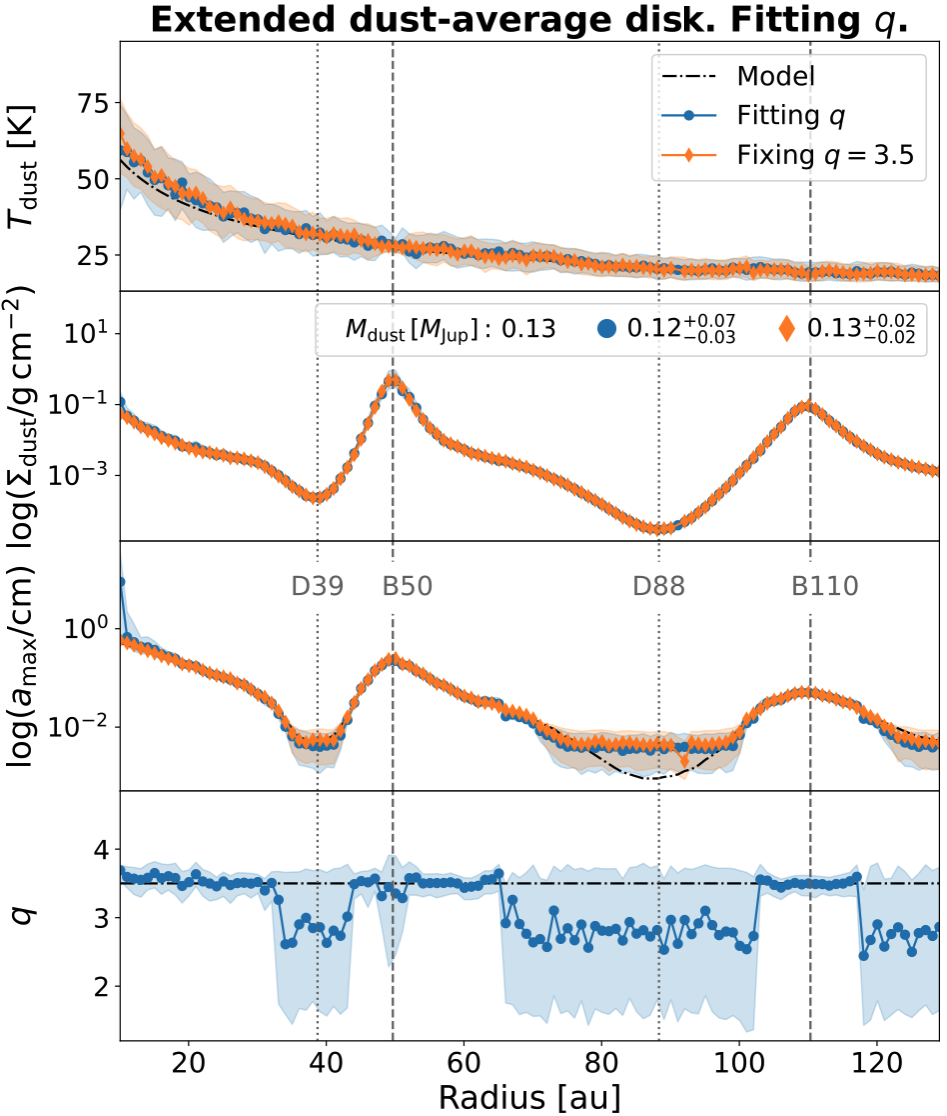}
   
   \caption{Same as Figure \ref{fig:p} but for the extended disk model (see Subsection \ref{ssec:bands}).}
    \label{fig:p2}%
\end{figure}

Figure \ref{fig:p} and \ref{fig:p2} show the dust temperature, dust surface density, maximum grain size and power law index of the grain size distribution (from \textit{top} to \textit{bottom}) measured from an SED analysis when fixing (\textit{orange} solution), or fitting (\textit{blue} solution) $q$. To not introduce limited resolution or optical depths effects in this comparison, we included in the SED intensity profiles from 0.87 mm to 7.46 mm with infinite resolution (see Subsection \ref{ssec:bands}). Figure \ref{fig:p} shows the comparison for the compact disk model, Figure \ref{fig:p2} for the extended disk model (see Section \ref{sec:data}). The dust temperature, dust surface density and maximum grain size measured in the two SED analyses are consistent with each other. Moreover, the \textit{bottom} panel of Figure \ref{fig:p} and \ref{fig:p2} suggests that this dust characterization method efficiently deals with the correlation between $a_\mathrm{max}$ and $q$. The only regions where we are not able to reproduce the expected value of $q$ are the marginally thick inner disk of the compact disk and the deep gaps of the extended disk. In these deep gaps, grains sizes are much smaller than the mean free path of the photons ($a < \lambda / 2\pi$). In this Rayleigh regime, the opacities are insensitive to the grain sizes and we are not able to constrain them from an SED analysis (see Figure 4 of \citealt{Birnstiel2018}).

\section{Linear vs. logarithmic sampling}
\label{app:sampling}
\begin{figure*}[h!]
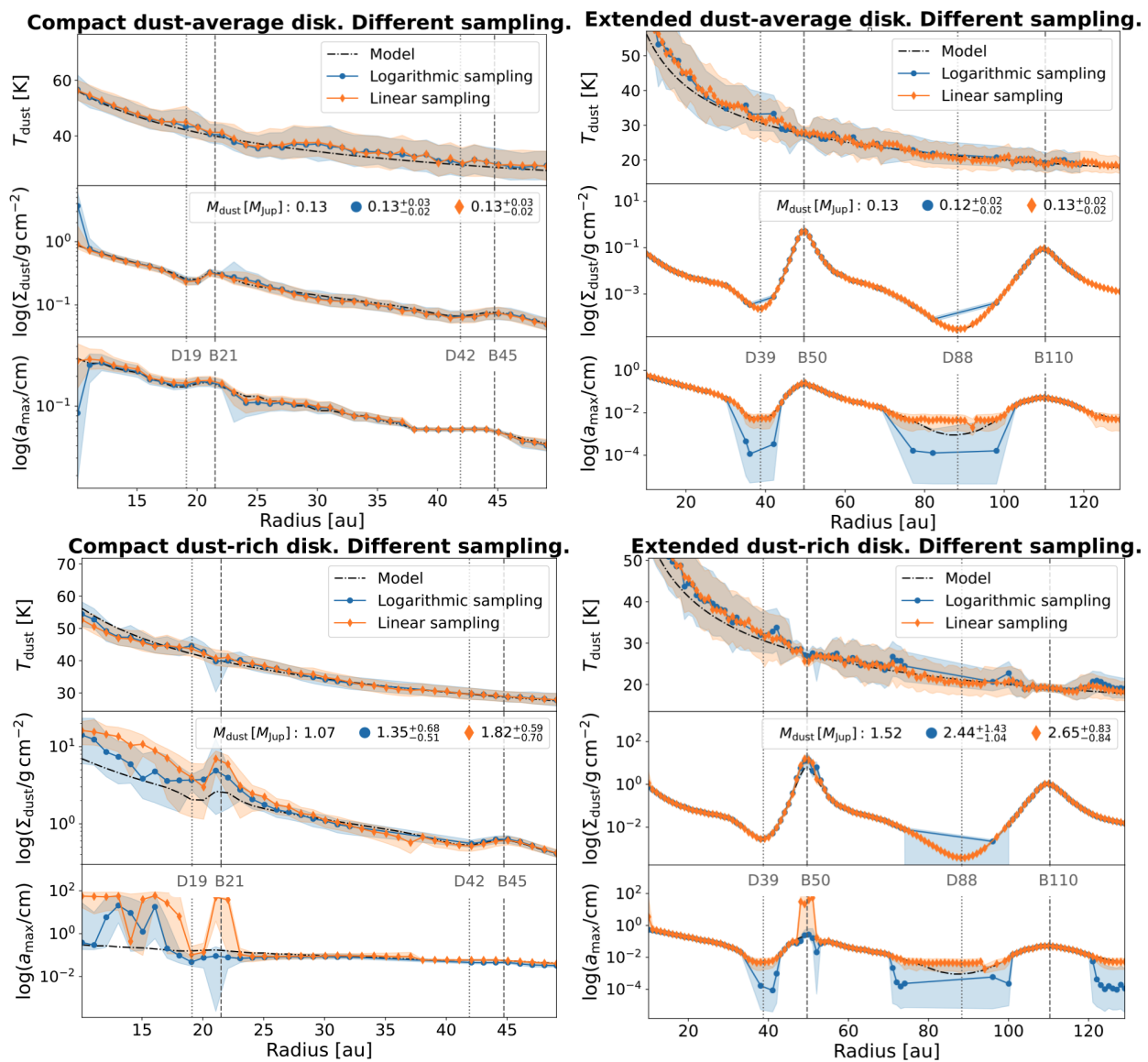
{}
   \centering
   \includegraphics[width=0.44\textwidth]{Final_Int_Sampling_Amaxmedian_posteriors.pdf}
   \includegraphics[width=0.44\textwidth]{Final_Int_Sampling_Amax2median_posteriors.pdf}   
    \includegraphics[width=0.44\textwidth]{Final_High_Sampling_Amaxmedian_posteriors.pdf}
    \includegraphics[width=0.44\textwidth]{Final_High_Sampling_Amax2median_posteriors.pdf}
   \caption{Dust temperature (\textit{top panel}), dust surface density (\textit{middle panel}), and maximum grain size (\textit{bottom panel}) measured from an SED analysis of the dust-average (\textit{top plots}) and dust-rich (\textit{bottom plots}) compact (\textit{left}) and extended (\textit{right}) disk models (see Subsection \ref{ssec:bands}). The SED includes emission from 0.87 mm to 7.46 mm, simulated with infinite resolution. The \textit{orange} solution is obtained by linearly sampling the parameter space, the \textit{blue} solution with a logarithmic sampling. The continuum line represents the 50th percentiles of the marginalized probability distribution of the parameters, the shaded area extends from the 16th to the 84th percentiles. The \textit{black} line indicates the dust properties of the models. The vertical gray lines represent the position of the bright (B) and dark (D) gaps in the dust surface density profile.}
    \label{fig:sampling}%
\end{figure*}
As discussed in Section \ref{sec:analysis}, in this work we estimate the dust properties of our disk models by comparing, through a Bayesian approach, their multi-band intensity profiles to a multi-wavelength 1D slab model (equation \ref{eq:inu}). In particular, we employ the code \texttt{emcee} \citep{ForemanMackey2019} to estimate the posterior distribution of the parameters of the model: the dust temperature, dust surface density, and maximum grain size. 

\texttt{Emcee} samples the N-dimensional parameter space, where N is the number of parameters, to find the N-tuple that best reproduces the data, while ensuring the exploration remains within the boundaries set by a prior. In this work we choose to define uniform priors and sample the parameter space in linear scale. 

For the maximum grain size, this might not be the most appropriate choice. Our choice of letting $a_\mathrm{max}$ vary between 0 cm and 100 cm does not actually reflect the grain size distribution we expect to detect at millimeter wavelengths. The peak of intensity at millimeter wavelengths is reached for grain sizes of the order of $\sim \lambda / 2\pi$, thus grains $\gtrsim 10$ cm emit at lower frequencies than the ones included in the SED. Furthermore, given the wide range of grain sizes expected in protoplanetary disks, spanning various orders of magnitude,  a uniform prior in the logarithmic space is a better representation of our physical expectations.

Figure \ref{fig:sampling} shows the dust temperature, dust surface density, and maximum grain size (from \textit{top} to \textit{bottom}) measured from SED analyses of the dust-rich (\textit{top plot}) and dust-average (\textit{bottom plot}) compact (\textit{left}) and extended (\textit{right}) disk models (see Subsection \ref{ssec:bands}). In each panel, two solutions are shown: in \textit{orange}, the results obtained if we sample $a_\mathrm{max}$ in linear space within 0 and 100 cm; in \textit{blue}, the solution obtained with a uniform prior in logarithmic space. The SED analyses include emission from 0.87 mm to 7.46 mm, simulated with infinite resolution. 

From the \textit{top plots} of Figure \ref{fig:sampling} it is clear that, if both optically thick and thin observations are included in the SED analyses, the dust properties of a disk are well reproduced independently from the sampling method. In fact, the two sampling methods yield different results only within the deep gaps of the extended disk model. These regions are optically thin across all wavelengths, and under the Mie theory of opacity, their dust opacity can be matched by two distinct dust populations (refer to Figure 4 of \citealt{Birnstiel2018}). Linear sampling tends to favor higher grain sizes, whereas logarithmic sampling identifies smaller grain sizes as a solution. While this serves as an interesting exercise, it is important to note that such regions, characterized by optically thin emission even at high frequencies (Band 7/6), are not commonly targeted by ALMA and are challenging to detect.

On the other hand, when the observations included in the SED analysis are not optically thin, we see inconsistency in the dust properties measured with the two sampling method (see \textit{bottom plots} of Figure \ref{fig:sampling}). The 7.46 mm emission of the dust-rich ($M_\mathrm{dust} \gtrsim 1.4 \,M_\mathrm{Jup}$) disk models are marginally thick in the inner disk and in the rings. In these dense regions, the dust properties measured by sampling the parameter space in linear (\textit{orange}) of logarithmic (\textit{blue}) scale differs. While neither of the SED analyses produce robustly constrained results, in a linear sampling the optical thickness of the observations is clearly highlighted by the \textit{walkers} converging toward the maximum of the prior. Instead, a uniform prior in logarithmic space favors lower $a_\mathrm{max}$ values, resulting in more reasonable solutions that are closer to the expected values but not necessarily accurate: in the inner disk of the compact disk (\textit{bottom left}), the measured maximum grain sizes can be $\sim 2$ order of magnitude larger than the true values.

To highlight difficult convergences and optical thickness, in our work we choose to linearly sample the parameter space. To test the stability of the solution of SED analyses of real observations, we suggest to double check the results with both a linear and logarithmic sampling or to explore the parameter space with other sampling techniques, such as a No-U-Turn Sampler (NUTS).

 \section{Flux calibration errors}
\label{app:flux}
Something that we need to consider while performing dust characterization analyses is that, on top of the uncertainties on the dust properties derived from an \texttt{emcee} analysis of a disk's SED, additional uncertainties arise from flux calibration errors that affect each multi-band intensity profile.
%the how the flux calibration error of each intensity profile included in the SED can affect the determination of the dust properties. 
\begin{figure}[h]{}
   \centering
   \includegraphics[width=\hsize]{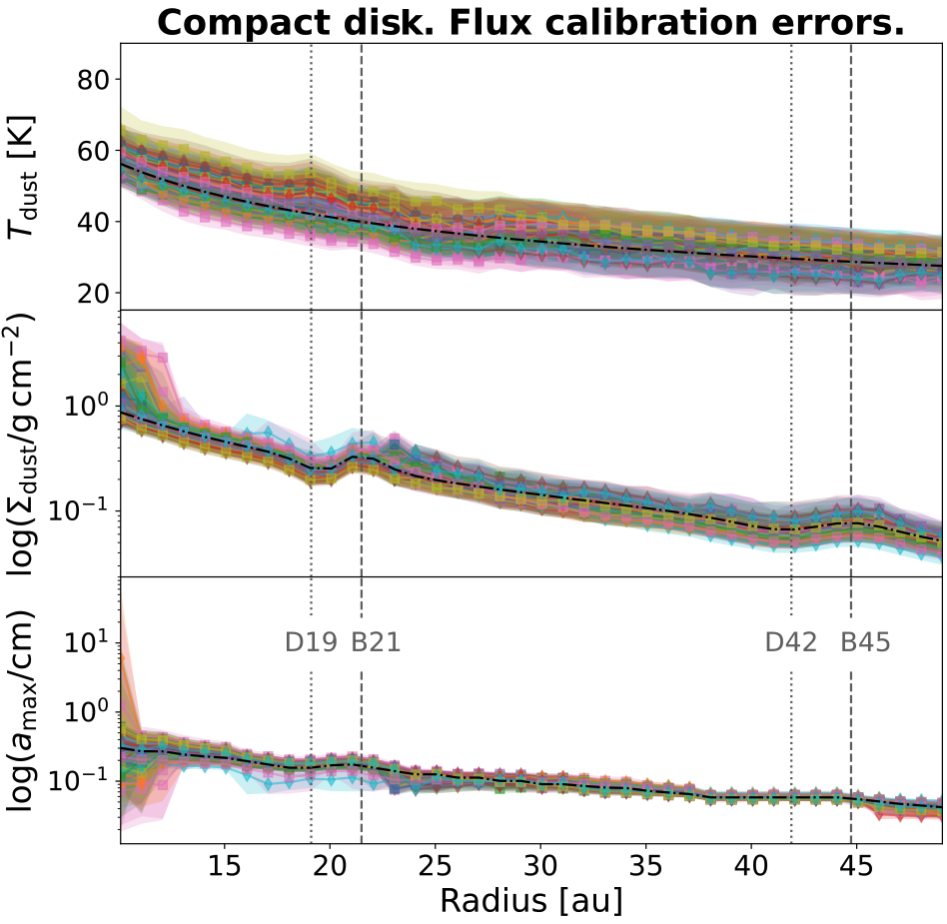}

   \caption{Dust temperature (\textit{top panel}), dust surface density (\textit{middle panel}), and maximum grain size (\textit{bottom panel}) measured from a SED analyses of compact disk model (see Subsection \ref{ssec:bands}). The SED includes emission from 0.87 mm to 7.46 mm, simulated with infinite resolution. In each panel, we are showing 50 different solutions obtained by rescaling each of the multi-band intensity profiles within typical flux calibration errors of radio facilities such as ALMA and VLA (10$\%$ at 0.87 mm, 1.29 mm, and 7.46 mm; 5$\%$ at 3.07 mm). The continuum line represents the 50th percentiles of the marginalized probability distribution of the parameters, the shaded area extends from the 16th to the 84th percentiles. The \textit{black} line indicates the dust properties of the models. The vertical gray lines represent the position of the bright (B) and dark (D) gaps in the dust surface density profile.}
    \label{fig:flux}%
\end{figure}

In the SED analyses of our simulated disk models, performed using the Ensemble Sampler \texttt{emcee} \citep{ForemanMackey2019}, we have already incorporated flux calibration errors. Indeed, we define the error on the multi-wavelength intensity profiles, and thus the 'allowed' distance between the model intensity and the data in the likelihood function, as the quadrature sum of two components: the standard error of the mean, obtained from azimuthally averaging the intensity profiles, and the flux calibration errors. We followed the guidelines from the ALMA Technical Handbook \citep{ALMA2023} and the Guide to Observing with the VLA \citep{VLA2023}, by setting flux calibration errors of $10\%$ at 0.87 mm, 1.29 mm, and 7.46 mm, and $5\%$ at 3 mm.

However, in our simulated multi-band disk observations, the fluxes are set to their expected values. Now, we directly estimate the spread introduced by flux calibration errors in the dust properties derived from the SEDs.

To do so, we perform 50 separate SED analyses. In each iteration, the intensity profiles at the different wavelengths are rescaled within their respective flux calibration error margins. The dust temperatures, dust surface densities and maximum grain sizes resulting from these 50 SED analyses including emission from 0.87 mm to 7.46 mm for the compact disk model are shown in Figure \ref{fig:flux}. 

By varying the multi-band fluxes within their calibration errors, we observe that the spread in dust temperature measurements is less than 17$\%$. Interestingly, the dust surface density and maximum grain size are also surprisingly well constrained. Except near the very inner disk, the order of magnitude for both of these parameters is consistently matched.
\end{appendix}

\end{document}